\documentstyle[epsf]{article}

\newcommand{\reseteqnum}{\setcounter{equation}{0}}

\newcommand{\bbone}{{\mathchoice {\rm 1\mskip-4mu l} {\rm 1\mskip-4mu l}
{\rm 1\mskip-4.5mu l} {\rm 1\mskip-5mu l}}}
\newcommand{\Tr}{\mbox{\rm Tr}}

\newcommand{\rvac}{\vert 0 \rangle}
\newcommand{\rvacp}{\vert + \rangle}
\newcommand{\rvacm}{\vert - \rangle}
\newcommand{\lvac}{\langle 0 }
\newcommand{\lvacp}{\langle + }
\newcommand{\lvacm}{\langle - }

\newcommand{\rVacp}{\vert v+ \rangle}
\newcommand{\rVacm}{\vert v- \rangle}
\newcommand{\lVacp}{\langle v+}
\newcommand{\lVacm}{\langle v-}

\title{
\hfill
\parbox{4cm}{\normalsize KUNS-1445\\HE(TH)~97/08}\\
\vspace{1cm}
Gauge Freedom in Chiral Gauge Theory \\
with Vacuum Overlap -- Two-dimensional case}
\author{Yoshio Kikukawa\thanks{e-mail address:
kikukawa@physics.rutgers.edu, kikukawa@gauge.scphys.kyoto-u.ac.jp}
\\
{\normalsize\em Department of Physics and Astronomy,
                Rutgers University\thanks{On leave of absence from:
         Department of Physics, Kyoto University, Kyoto 606-01, Japan}
}\\
{\normalsize\em Piscataway, NJ 08855-0845}
}

\begin{document}
\maketitle

\begin{abstract}
Dynamical nature of the gauge degree of freedom and its effect to
fermion spectrum are studied at $\beta=\infty$ for two-dimensional
nonabelian chiral gauge theory in the vacuum overlap formulation.
It is argued that the disordered gauge degree of freedom does not
necessarily cause the massless chiral state in the (waveguide)
boundary correlation function.
An asymptotically free self-coupling for the gauge degree of freedom
is introduced by hand at first. This allows us to tame the gauge
fluctuation by approaching the critical point of the gauge degree
of freedom without spoiling its disordered nature.
We examine the spectrum in the boundary correlation function and
find the mass gap of the order of the lattice cutoff.
There is no symmetry against it. Then we argue that the decoupling
of the gauge freedom can occur as the self-coupling is removed,
provided that the IR fixed point due to the Wess-Zumino term is
absent by anomaly cancellation.
\end{abstract}

\newpage
\section{Introduction }
\reseteqnum
It has been one of the most important issues to clarify
the dynamical behavior of the gauge freedom and its effect to
the fermion spectrum for various proposals of lattice
chiral gauge theory\cite{
wilson-yukawa-model,
aoki-kashiwa-funakubo,
wilson-yukawa-model-analysis,
eichten-preskill,
eichten-preskill-analysis,
staggered-fermion,
staggered-fermion-analysis,
original-kaplan,
original-kaplan-analysis,
original-waveguide,
waveguide-analysis-golterman,
waveguide-analysis-general,
waveguide-four-fermi,
waveguide-majorana-yukawa,
original-overlap,overlap}.
In this article, we discuss this issue in the context of
the vacuum overlap formulation\cite{original-overlap}.

In the vacuum overlap formulation of a generic chiral gauge theory,
gauge symmetry is explicitly broken by the complex phase of
fermion determinant. In order to restore the gauge invariance,
gauge average ---the integration along gauge orbit--- is invoked.
Then, what is required for the dynamical nature of the
gauge freedom at $\beta=\infty$ (pure gauge limit) is that
the global gauge symmetry is not broken spontaneously and
all the bosonic field of the gauge freedom could be heavy compared
to a typical mass scale of the theory so as to decouple from physical
spectrum\cite{gauge-symmetry-restoration,wilson-yukawa-model,
aoki-kashiwa-funakubo,staggered-fermion,original-overlap}.

However, through the analysis of the waveguide
model\cite{original-waveguide}, it has been claimed
that this required disordered nature of the gauge freedom
causes the vector-like spectrum of
fermion\cite{waveguide-analysis-golterman}.
In this argument, the fermion correlation functions at
the waveguide boundaries were examined.
One may think of
the counter parts of these correlation functions in the overlap
formulation by putting creation and annihilation operators in
the overlap of vacua with the same signature of mass.
Let us refer this kind of correlation function as
{\it boundary correlation function} and
the correlation function in the original definition as
{\it overlap correlation function}.
We should note that the boundary correlation functions
are no more the observables in the sense defined in the overlap
formulation\cite{original-overlap}; they cannot be expressed by
the overlap of two vacua with their phases fixed by the Wigner-Brillouin
phase convention. But, they are still relevant because they can probe
the auxiliary fermionic system for the definition of
the complex phase of chiral determinant and therefore the
anomaly (the Wess-Zumino term).
If massless chiral states could actually appear in the boundary
correlation functions, we would have difficulty defining the
complex phase.

Our objective in this paper is to argue that
the disordered nature of the gauge degree of freedom
can maintain
the chiral spectrum of these correlation functions.
For this goal, we concentrate on the pure gauge limit of
a two-dimensional $SU(2)$ nonablelian chiral gauge theory.
Then it is naturally expected that the gauge freedom
acquires mass dynamically, because of its two-dimensional and
nonabelian nature. To make it explicit,
an asymptotically free self-coupling for the gauge freedom
is introduced by hand at first.\footnote{This gauge breaking
term is usually induced from fermion determinant.
In the overlap formulation, however, it is absent because
the gauge symmetry breaking occurs only in the complex
phase of the fermion determinant. }
This asymptotic freedom allows us to tame the gauge fluctuation
by approaching the critical point for the gauge freedom
without spoiling its disordered nature.
There we can show by the spin wave approximation that
the spectrum in the boundary correlation function
has mass gap of the order of the lattice cutoff
and it survives the quantum correction due to the gauge fluctuation.
Since the overlap correlation function does not depend on the
gauge freedom and does show the chiral
spectrum\cite{original-overlap},
the above fact means that the entire fermion spectrum is chiral.
Then, we argue that how far the mass of the gauge freedom can be lifted
without affecting the chiral spectrum as the self-coupling is removed.
In this respect, we will note that there is no symmetry
against the spectrum mass gap of the boundary correlation function.
We will also note that the absence of the IR fixed point due to
the Wess-Zumino term is crucial for the decoupling of the gauge freedom.
Finally, the effect of the (waveguide) boundary Yukawa coupling is
also clarified in the context of the vacuum overlap, using the
operator technique\cite{boundary-yukawa-coupling-operator}
to incorporate the Yukawa coupling into overlap.

A few comments are in order.
Note that our result is consistent with what was found
in the Wilson-Yukawa model. In this model,
the so-called strong coupling symmetric (PMS) phase has been
identified as the phase which can fulfill the two
requirement: the disordered gauge freedom and the chiral spectrum of
fermion.\footnote{
Actual reasons why the Wilson-Yukawa formulation is not considered to
be able to describe the chiral gauge theory are the triviality of
the chiral coupling to vector bosons in four-dimensions
and the fermion number conservation\cite{wilson-yukawa-model-analysis}.}
In fact, the second requirement is subtle.  But if the
shift symmetry\cite{shift-symmetry} is invoked it is also
fulfilled effectively because one of the {\it Weyl} components
of the massless {\it Dirac} fermion can be made decoupled from
any other particles.

For the case of the two-dimensional $SU(2)$ chiral gauge theory,
the pure gauge dynamics in the Wilson-Yukawa model is nothing
but the dynamics of the two-dimensional chiral $SU(2)$ nonlinear
sigma model. There appears only the paramagnetic phase.
The fermion spectrum can be easily examined at the critical point
by virtue of the asymptotic freedom.
The fermion mass does not vanish at the critical point and
it is found that the phase is actually the PMS phase.
This is the very strategy with which
we will argue in the context of the vacuum overlap formulation.

The possible existence of the PMS phase in the waveguide formulation
was already argued in the recent proposal of the gauge invariant
formulation of the standard model with the invariant
four-fermion operators\cite{waveguide-four-fermi}.
Quite recently, it was also argued in the system
with the Majorana type Yukawa coupling\cite{waveguide-majorana-yukawa}.
Our result is suggesting that,
in the case of the two-dimensional nonabelian gauge theory,
the PMS phase can also exist in the waveguide model
with the Dirac type Yukawa coupling.

This paper is organized as follows.
In section~\ref{sc:pure-gauge-limit}, we first define with
vacuum overlap a two-dimensional $SU(2)$ chiral gauge theory.
Then, we clarify its structure in the pure gauge limit.
As observables in the pure gauge limit, we define
the boundary correlation functions.
In section~\ref{sc:pure-gauge-dynamics},
we calculate the induced imaginary action
(including the Wess-Zumino term) and discuss the
dynamical nature of the gauge freedom.
In section~\ref{sc:boundary-fermion-correlation-function},
the boundary correlation functions are
examined near the critical point.
In section~\ref{sc:decoupling-of-gauge-freedom},
we discuss the possible decoupling of the gauge freedom.
In section~\ref{sc:boundary-Yukawa-coupling},
the effect of the boundary Yukawa coupling is examined.
In section~\ref{sc:conclusion-discussion},
we summarize and discuss our result
from the point of view of the Nielsen-Ninomiya theorem
and its extension for the case with interaction.

\section{Pure Gauge Limit }
\reseteqnum

\label{sc:pure-gauge-limit}
\subsection{Two-dimensional $SU(2)$ Chiral Gauge Theory}

To be specific, we consider the two-dimensional $SU(2)$ gauge
theory with four $SU(2)$ doublets of left-handed Weyl fermions
and one $SU(2)$ Adjoint of right-handed Weyl
fermions\cite{massless-boundstate-fermion-2d}.
This representation of fermion is anomaly free.
\begin{equation}
  \sum_r k_r = 4 \times 1 - 4 = 0 ,
\end{equation}
where
\begin{equation}
\Tr \left( T^a T^b \right)_r = \delta^{ab} \frac{1}{2} k_r  .
\end{equation}

The partition function of the chiral gauge theory
is given by the following formula in the vacuum overlap formulation.
\begin{equation}
    Z= \int [dU] \exp \left(-\beta S_G \right)
      \prod_{rep.} \left(
        \frac{\lvacp \rVacp}{|\lvacp\rVacp|}
               \lVacp \rVacm
        \frac{ \lVacm \rvacm}{| \lVacm \rvacm|}
              \right) .
\end{equation}
In this formula,
$\rVacp$ and $\rVacm$ are the vacua of the second-quantized
Hamiltonians of the three-dimensional Wilson fermion
with positive and negative bare masses, respectively.
\begin{equation}
  \hat H_{\pm} = \hat a^\dagger_{n\alpha}{}^i
  H_{\pm \, n\alpha,m\beta}{}_i^j \hat a_{m\beta}{}_j ,
\end{equation}

\begin{eqnarray}
H_{\pm\, n\alpha,m\beta}{}_i^j
&=&
\left( \begin{array}{cc}
B_{nm}{}_i^j \pm m_0 \delta_{nm} \delta_i^j
& C_{nm}{}_i^j \\
C^\dagger_{nm}{}_i^j
& -B_{nm}{}_i^j \mp m_0 \delta_{nm} \delta_i^j
\end{array} \right) ,
\\
B_{nm}{}_i^j
&=& \frac{1}{2}\sum_\mu \left( 2 \delta_{n,m}\delta_i^j
-\delta_{n+\hat\mu,m}U_{n\mu}{}_i^j
-\delta_{n,m+\hat\mu}U^\dagger_{m\mu}{}_i^j
\right) , \\
C_{nm}{}_i^j &=& \frac{1}{2} \sum_\mu \sigma_\mu
\left(
\delta_{n+\hat\mu,m}U_{n\mu}{}_i^j
 -\delta_{n,m+\hat\mu}U^\dagger_{m\mu}{}_i^j
\right),  \quad \sigma_\mu= ( i, 1 ).
\end{eqnarray}
$\rvacp$ and $\rvacm$ are corresponding free vacua.
The Wigner-Brillouin phase convention is explicitly implemented
by the overlaps of vacua with the same signature of mass.
$\prod_{rep.}$ stands for the product over all Weyl fermion multiplets
in the anomaly free representation. $S_G$ is the gauge action.

\subsection{Pure gauge limit $\beta = \infty$}

In the vanishing gauge coupling limit $\beta =\infty$,
the gauge link variable is given in the pure gauge form:
\begin{equation}
  U_{n\mu}= g_n g_{n+\mu}^\dagger \quad  g^{}_n \in SU(2) .
\end{equation}
Then the model describes the gauge degree of freedom
coupled to fermion through gauge non-invariant piece of complex
phase of chiral determinants.
\begin{eqnarray}
    Z&=& \int [dg]
      \prod_{rep.} \left(
   \frac{\lvacp \vert \hat G \rvacp}{|\lvacp \vert \hat G \rvacp|}
   \lvacp \rvacm
   \frac{\lvacm \vert \hat G^\dagger \rvacm}
        {|\lvacm \vert \hat G^\dagger \rvacm|}
              \right) .
\end{eqnarray}
$G$ is the operator of the gauge transformation given by:
\begin{equation}
\hat G
= \exp \left( \hat a_n^{\dagger i} \{\log g\}_i{}^j \hat a_{n j} \right) .
\end{equation}

To control the fluctuation of the gauge degree of freedom, we add
the following gauge non-invariant term to the original
model\cite{gauge-symmetry-restoration,
           staggered-fermion,staggered-fermion-analysis}:
\begin{equation}
\exp \left\{ K \sum_{n\mu} {\rm Tr}( U_{n\mu} + U_{n\mu}^\dagger)
\right\},    \qquad K \equiv \frac{1}{\lambda^2} .
\end{equation}
Then, in the pure gauge limit, the model reduces to just the
two-dimensional chiral $SU(2)$ nonlinear sigma model coupled to anomaly
free chiral fermions through the gauge non-invariant piece of complex
phase of chiral determinants.
\begin{eqnarray}
\label{eq:Chiral-pure-gauge-partition-function-main}
    Z&=& \int [dg]
\exp \left\{
K \sum_{n\mu} {\rm Tr}( g_n g_{n+\mu}^\dagger  + g_{n+\mu}g_n^\dagger )
     \right\}
      \prod_{rep.} \left(
   \frac{\lvacp \vert \hat G \rvacp}{|\lvacp \vert \hat G \rvacp|}
   \lvacp \rvacm
   \frac{\lvacm \vert \hat G^\dagger \rvacm}
        {|\lvacm \vert \hat G^\dagger \rvacm|}
              \right)
\nonumber\\
&\equiv& \int d\mu[g;K] .
\end{eqnarray}
Including the complex action,
the functional integral measure of the gauge freedom is denoted
by $d\mu[g;K]$.

This model is invariant under two global $SU(2)$ transformations
acting on the chiral field. The first one is the global remnant
of gauge transformation:
\begin{equation}
 SU(2)_{global}: \qquad   g_n{}_i^j \longrightarrow g_0{}_i^k g_n{}_k^j .
\end{equation}
The second one comes from the arbitrariness of choice of pure gauge
variable $g_n$:
\begin{equation}
 SU(2)_{hidden}: \qquad
g_n{}_i^j \longrightarrow g_n{}_i^k h^\dagger{}_k^j .
\end{equation}
They defines the chiral transformation of
$G= SU(2)_L \times SU(2)_R = SU(2)_{global} \times SU(2)_{hidden}$
and the model is symmetric under this chiral transformation.

We refer
the imaginary action of the gauge freedom
induced from the fermion determinant
as the null Wess-Zumino action
because the actual Wess-Zumino terms are canceled among the fermions.
We denote it by $\Delta\Gamma_{WZ}$:
\begin{equation}
  e^{i \Delta\Gamma_{WZ}[g]} \equiv
   \prod_{rep.} \left(
   \frac{\lvacp \vert \hat G \rvacp}
  {|\lvacp \vert \hat G \rvacp|}
  \lvacp \rvacm
   \frac{\lvacm \vert \hat G^\dagger \rvacm}
        {|\lvacm \vert \hat G^\dagger \rvacm|}
              \right) .
\end{equation}

\subsection{Observables in pure gauge limit}

In order to probe the fermion spectrum in the pure gauge limit,
we examine fermion correlation functions defined by
putting creation and annihilation operators in
the overlaps of two vacua with the same signature of masses,
but interacting and free. We refer this kind of correlation function
as {\it boundary correlation function}. The fermion correlation
function in the original definition\cite{original-overlap} is refered
as {\it overlap correlation function}.

\subsubsection{Boundary fermion correlation function}
As for the overlap of the vacuum of the Hamiltonian with negative mass,
there are three possible definitions of boundary correlation function
in the representation $r$:
\begin{eqnarray}
\langle \phi_n{}_i \phi^\dagger_m{}^j  \rangle_{-r}
&\equiv&\frac{1}{Z} \int d\mu[g;K]
\frac{\lvacm \vert \hat G^\dagger
  \left\{ \hat a_n{}_i \hat a^\dagger_m{}^j
         -\frac{1}{2} \delta_{nm} \delta_i^j \right\}
      \rvacm_r}
     {\lvacm \vert \hat G^\dagger \rvacm_r } ,
\\
\langle \varphi_n{}_i \varphi^\dagger_m{}^j  \rangle_{-r}
&\equiv&\frac{1}{Z} \int d\mu[g;K] \,
\frac{\lvacm \vert
  \left\{ \hat a_n{}_i \hat a^\dagger_m{}^j
         -\frac{1}{2} \delta_{nm} \delta_i^j \right\}
      \hat G^\dagger \rvacm_r}
     {\lvacm \vert \hat G^\dagger \rvacm_r } ,
\\
\label{eq:boundary-correlation-function-gauge-mixed}
\langle \varphi_n{}_i \phi^\dagger_m{}^j  \rangle_{-r}
&\equiv&\frac{1}{Z}\int d\mu[g;K] \,
\frac{\lvacm \vert
  \left\{ \hat a_n{}_i \hat G^\dagger \hat a^\dagger_m{}^j
         -\frac{1}{2} \delta_{nm} \left(g^\dagger_m\right)_i^j \right\}
      \rvacm_r}
     {\lvacm \vert \hat G^\dagger \rvacm_r } .
\nonumber\\
\end{eqnarray}

The transformation properties of these correlation functions
under the chiral $SU(2)$ can be read as follows:
\begin{equation}
\langle \phi_n{}_i \phi^\dagger_m{}^j  \rangle_{-r}
\longrightarrow
(g_0{}_i^s)
\langle \phi_n{}_s \phi^\dagger_m{}^t  \rangle_{-r}
(g_0^\dagger{}_t^j)   ,
\end{equation}
\begin{equation}
\langle \varphi_n{}_i \varphi^\dagger_m{}^j  \rangle_{-r}
\longrightarrow
(h_i^s)
\langle \varphi_n{}_s \varphi^\dagger_m{}^t  \rangle_{-r}
(h^\dagger{}_t^j)   ,
\end{equation}
\begin{equation}
\langle \varphi_n{}_i \phi^\dagger_m{}^j  \rangle_{-r}
\longrightarrow
(h_i^s)
\langle \varphi_n{}_s \phi^\dagger_m{}^t  \rangle_{-r}
(g_0^\dagger{}_t^j)   .
\end{equation}
As we will argue in the next section, the pure gauge model
we are considering has the same infrared structure as
the chiral $SU(2)$ nonlinear sigma model.
Then, the local order parameter
$ \langle \left( g_m \right)_i^j \rangle $ is not well-defined
and only chiral $SU(2)$ invariant quantities can be used as
observables\cite{observable-in-nonlinear-sigma-model}.
The first and second correlation functions can be made invariant under
the chiral $SU(2)$ by taking the trace over the group indices.
The third one cannot be made invariant
and we should discard it from observables.

As for the overlap of the vacuum of the Hamiltonian with positive mass,
there are also three possible definitions of boundary correlation
function in the representation $r$:
\begin{eqnarray}
\langle \phi_n{}_i \phi^\dagger_m{}^j  \rangle_{+r}
&\equiv&\frac{1}{Z} \int d\mu[g;K]
\frac{\lvacp \vert
  \left\{ \hat a_n{}_i \hat a^\dagger_m{}^j
         -\frac{1}{2} \delta_{nm} \delta_i^j \right\}
      \hat G \rvacp_r}
     {\lvacp \vert \hat G \rvacp_r } ,
\\
\langle \varphi_n{}_i \varphi^\dagger_m{}^j  \rangle_{+r}
&\equiv&\frac{1}{Z} \int d\mu[g;K] \,
\frac{\lvacp \vert \hat G
  \left\{ \hat a_n{}_i \hat a^\dagger_m{}^j
         -\frac{1}{2} \delta_{nm} \delta_i^j \right\}
      \rvacp_r}
     {\lvacp \vert \hat G \rvacp_r } ,
\\
\langle \phi_n{}_i \varphi^\dagger_m{}^j  \rangle_{+r}
&\equiv&\frac{1}{Z}\int d\mu[g;K] \,
\frac{\lvacp \vert
  \left\{ \hat a_n{}_i \hat G \hat a^\dagger_m{}^j
         -\frac{1}{2} \delta_{nm} \left(g^\dagger_m\right)_i^j \right\}
      \rvacp_r}
     {\lvacp \vert \hat G \rvacp_r } .
\nonumber\\
\end{eqnarray}

They transform under the chiral $SU(2)$ as follows:
\begin{equation}
\langle \phi_n{}_i \phi^\dagger_m{}^j  \rangle_{+r}
\longrightarrow
(g_0{}_i^s)
\langle \phi_n{}_s \phi^\dagger_m{}^t  \rangle_{+r}
(g_0^\dagger{}_t^j)   ,
\end{equation}
\begin{equation}
\langle \varphi_n{}_i \varphi^\dagger_m{}^j  \rangle_{+r}
\longrightarrow
(h_i^s)
\langle \varphi_n{}_s \varphi^\dagger_m{}^t  \rangle_{+r}
(h^\dagger{}_t^j)   ,
\end{equation}
\begin{equation}
\langle \phi_n{}_i \varphi^\dagger_m{}^j  \rangle_{+r}
\longrightarrow
(g_0{}_i^s)
\langle \phi_n{}_s \varphi^\dagger_m{}^t  \rangle_{+r}
(h^\dagger{}_t^j) .
\end{equation}
The first two correlation functions can be made invariant
under the chiral $SU(2)$ by taking the trace over the group indices.
But the third one cannot be made invariant. We discard this
correlation function from the observables in the pure gauge limit.

Note that contrary to the case of the overlap correlation functions,
the gauge freedom does not decouple from these boundary correlation
functions.
In the section~\ref{sc:boundary-fermion-correlation-function},
we will examine the two invariant correlation functions associated to
the overlap of the vacuum of the Hamiltonian with negative mass.
The positive mass case could be examined in a similar manner.

\section{Pure Gauge Dynamics}
\reseteqnum

\label{sc:pure-gauge-dynamics}
Let us first discuss the dynamics of the gauge freedom without
the null Wess-Zumino action. Then the model reduces to the
chiral $SU(2)$ nonlinear sigma model.
The dynamics of this model is well-known.
First of all, the coupling constant $\lambda$
is asymptotically free. This suggests that the model
develops the mass gap dynamically and it is actually the case.
The chiral $SU(2)$ symmetry is realized linearly
in the entire region of the coupling constant $\lambda$,
in accord with the general theorem in
two-dimensions.\cite{mermin-wagner-coleman}

In order to examine the effect of the null Wess-Zumino action
on the dynamics of the chiral $SU(2)$ nonlinear sigma model,
we will evaluate the null Wess-Zumino action in perturbation
theory of $\lambda$.
The explicit formula of the contribution to the null Wess-Zumino action
in the representation $r$ reads
\begin{eqnarray}
i\Delta\Gamma_{WZ}[g]_r
&=&
i {\rm Im} {\rm Tr} \, {\rm Ln} \left[
\sum_{m}
v^\dagger_-(p) e^{-i p m}
\left( g_m^\dagger{}_i^j \right)  e^{i q m } v_-(q)
\right]
\nonumber\\
&+&
i {\rm Im} {\rm Tr} \, {\rm Ln} \left[
\sum_{m}
v^\dagger_+(p) e^{-i p m}
\left( g_m{}_i^j \right)  e^{i q m }  v_+(q)
\right] .
\nonumber\\
\end{eqnarray}
The gauge freedom is expanded as
\begin{equation}
g_m^\dagger{}_i^j
= \bbone{}_i^j -i  \lambda \pi{}_i^j
  +(-i)^2 \frac{1}{2!} \lambda^2 \pi^2{}_i^j
  +(-i)^3 \frac{1}{3!} \lambda^3 \pi^3{}_i^j + \cdots .
\end{equation}
Then, the first nontrivial term in the negative mass contribution
turns out to be ${\cal O}(\pi^3)$ and it is evaluated as follows:
\begin{eqnarray}
&&
i \Delta\Gamma_{WZ}[\pi]_{r-}
\nonumber\\
&&
= i \frac{1}{3} \lambda^3
\int \frac{d^2 p_1}{(2\pi)^2}
\frac{d^2 p_2}{(2\pi)^2} \frac{d^2 p_3}{(2\pi)^2}
(2\pi)^2 \delta^2(p_1+p_2+p_3)
i \epsilon^{abc} \left( \pi^a(p_1) \pi^b(p_2) \pi^c(p_3) \right)
\times
\nonumber\\
&& \qquad \qquad
A_r
\int \frac{d^2 q}{(2\pi)^2}
{\rm Tr} \left[ S^v_-(q+p_1)S^v_-(q)S^v_-(q-p_2) \right] + \cdots ,
\end{eqnarray}
where
\begin{equation}
  S^v_-(p)\equiv \frac{1}{v^\dagger_-(p)v_-(p)}
   v_-(p) v^\dagger_-(p) ,
\end{equation}
\begin{equation}
{\rm Tr}\left( T^a T^b T^c \right)_r = i \epsilon^{abc} A_r  .
\end{equation}
See the appendix~\ref{appendix:null-Wess-Zumino-action} for the detail
of the calculation. For the anomaly free theory in consideration,
\begin{equation}
  \sum_r A_r = 4 \times \frac{1}{4} - 1 = 0 .
\end{equation}
The positive mass contribution also vanishes up to ${\cal O}(\pi^3)$.
This means that the induced null Wess-Zumino action vanishes completely
up to ${\cal O}(\pi^3)$.  Note that this is true before the expansion
with respect to external momentum in order to take the continuum limit.

The contribution from each fermion to the null Wess-Zumino action
should contain the Wess-Zumino term in the continuum limit.
\begin{eqnarray}
\int \frac{d^2 q}{(2\pi)^2}
{\rm Tr} \left[ S^v_-(q+p_1)S^v_-(q)S^v_-(q-p_2) \right]
&=& - \epsilon_{\mu\nu} p_{1\mu} p_{2\nu} c_{WZ}{}_-  + \cdots .
\nonumber\\
\end{eqnarray}
The coefficient $c_{WZ}{}_-$ is evaluated following the
technique of \cite{chern-simons-current} as follows:
\begin{eqnarray}
&&
c_{WZ}{}_-
\nonumber\\
&&=
\int \frac{d^2 q}{(2\pi)^2}
\frac{1}{2!} \epsilon_{\mu\nu}
{\rm Tr} \left[ \partial_\mu S^v_-(q) S^v_-(q) \partial_\nu S^v_-(q)
\right]
\nonumber\\
&&= (-i) \frac{1}{2}
\int \frac{d^2 q}{(2\pi)^2} \int_{-\infty}^\infty \frac{d \omega}{(2\pi)}
\nonumber\\
&& \quad
\frac{1}{2!} \epsilon_{\mu\nu}
{\rm Tr} \left\{ S_3(q,\omega) \partial_\mu S_3^{-1}(q,\omega)
                 S_3(q,\omega) \partial_\nu S_3^{-1}(q,\omega)
                 S_3(q,\omega) \partial_\omega S_3^{-1}(q,\omega)
         \right\}
\nonumber\\
&&=  (-i) \frac{1}{2}  \frac{1}{4\pi} \sum_{k=0}^{2} (-1)^k
      \left(
      \begin{array}{c} 2 \\ k \end{array} \right)
      \frac{ m_k } { \vert m_k \vert }
\qquad m_k= \left\{
  \begin{array}{c} -m_0  \quad (k=0) \\
                   (2k-m_0) \quad (k\not=0) \end{array} \right.
\nonumber\\
&&=(-i) \frac{1}{4\pi}  .
\end{eqnarray}
where
\begin{eqnarray}
S_3^{-1}(p,\omega)
&\equiv &
i \gamma_5 \omega + i \gamma_\mu C_\mu(p) + B-m_0 .
\nonumber\\
\end{eqnarray}
See appendix \ref{appendix:null-Wess-Zumino-action} for the detail
of the calculation. Similarly, we obtain for the positive mass
\begin{equation}
c_{WZ}{}_+ = 0 .
\end{equation}
Then we can see that they reproduce the correct value of the
continuum theory\cite{wess-zumino-term-overlap}.

Since the null Wess-Zumino action vanishes up to ${\cal O}(\pi^3)$,
it does not contribute to the quantum correction of
the action of the chiral $SU(2)$ nonlinear sigma model at the
one-loop order.
(The calculation is best performed by background field method.)
Then the beta function of $\lambda$ remains same as that of
the pure chiral $SU(2)$ nonlinear sigma model at one-loop order:
\begin{equation}
- a \frac{d \lambda^2}{d a }
= - \frac{\lambda^4}{4\pi} .
\end{equation}
Thus, the null Wess-Zumino action does not affect the renormalization
group properties of the nonlinear sigma model in the vicinity
of the critical point.
$\lambda$ remains asymptotically free.

The asymptotic freedom suggests that
the gauge freedom in the pure gauge limit
develops the mass gap dynamically,
even with the imaginary null Wess-Zumino action.
Accordingly, the chiral $SU(2)$ symmetry is realized linearly.
The scale of the mass gap is set by the coupling $\lambda$
through the renormalization group invariant scale.

\section{Boundary fermion on and off the critical point}
\reseteqnum

\label{sc:boundary-fermion-correlation-function}
$\lambda=0$ is the critical point of the pure gauge
model. At this point, the fluctuation of the gauge degree of freedom
is reduced completely to the global degree of freedom,
\begin{equation}
  g_n \longrightarrow g_0, \qquad \lambda \rightarrow 0,
\end{equation}
which decouples because of $SU(2)_{global}$ symmetry.
Then the spectrum of the boundary correlation functions can be
examined exactly. We will find that they have a mass gap of the
order of the lattice cutoff in the spectrum.

Leaving off the critical point,
the perturbation theory in $\lambda$ (spin wave approximation)
can be a good approximation due to asymptotic freedom,
as long as short distance quantities are concerned.
Since the boundary correlation functions have actually
the short distance nature, we can evaluate the quantum correction to
them by the perturbation theory (spin wave approximation).
We will find that the short distance nature persists inside the
symmetric phase off the critical point.

\subsection{Expression of boundary correlation functions}
The invariant boundary correlation functions associated
to the overlap of the vacuum of the Hamiltonian with negative mass
are evaluated as follows:
\begin{eqnarray}
\langle \phi_n{}_i \phi^\dagger_m{}^i  \rangle_{-r}
&\equiv&\frac{1}{Z} \int d\mu[g;K]
\frac{\lvacm \vert \hat G^\dagger
  \left\{ \hat a_n{}_i \hat a^\dagger_m{}^i
         -\frac{1}{2} \delta_{nm} \delta_i^i \right\}
      \rvacm_r}
     {\lvacm \vert \hat G^\dagger \rvacm_r }
\nonumber\\
&=&\frac{1}{Z} \int d\mu[g;K]
\left[
\frac{1}{2} \delta_{nm} \delta_i^i
- S_-^v[g](n;m){}_i^o \left( g^\dagger_m{}_o^i \right)
\right] ,
\\
\langle \varphi_n{}_i \varphi^\dagger_m{}^i  \rangle_{-r}
&\equiv&\frac{1}{Z} \int d\mu[g;K] \,
\frac{\lvacm \vert
  \left\{ \hat a_n{}_i \hat a^\dagger_m{}^i
         -\frac{1}{2} \delta_{nm} \delta_i^i \right\}
      \hat G^\dagger \rvacm_r}
     {\lvacm \vert \hat G^\dagger \rvacm_r }
\nonumber\\
&=&\frac{1}{Z} \int d\mu[g;K]
\left[
\frac{1}{2} \delta_{nm} \delta_i^i
- \left( g^\dagger_m{}_i^o \right) S_-^v[g](n;m){}_o^i
\right] ,
\end{eqnarray}
where
\begin{eqnarray}
S_-^v[g](n,m){}_i^j
&\equiv&
\int \frac{d^2 p}{(2\pi)^2} \frac{d^2 q}{(2\pi)^2} \times
\nonumber\\
&&
e^{i p n } v_-(p)
\left[ v^\dagger_-(q)
e^{-i q r} \left( g_r^\dagger{}_j^i \right) e^{i p r }
v_-(p) \right]^{-1}_{(p,i)(q,j)}
v^\dagger_-(q) e^{-i q m} .
\nonumber\\
\end{eqnarray}
See appendix~\ref{appendix:boundary-correlation-function}
for the detail of the calculation.

\subsection{Boundary correlation functions at criticality}

At the critical point, $\lambda=0$, the correlation functions
reduce to the expression:
\begin{eqnarray}
\langle \phi_n{}_i \phi^\dagger_m{}^i  \rangle_{-r}
&=&
\langle \varphi_n{}_i \varphi^\dagger_m{}^i  \rangle_{-r} ,
\\
\langle \varphi_n{}_i \varphi^\dagger_m{}^i  \rangle_{-r}
&=&
\int \frac{d^2 p}{(2\pi)^2} e^{i p (n-m)} \delta_i^i \times
\nonumber\\
&& \quad
\frac{1}{2 \lambda_-}
\left( \begin{array}{cc}
-m_0+B(p) & C(p) \\
C^\dagger(p) & m_0-B(p)
       \end{array}\right) ,
\end{eqnarray}
where
\begin{equation}
\lambda_- = {\sqrt{C_\mu^2(p)+(B(p)-m_0)^2}} .
\end{equation}

This boundary correlation function does not show any pole which
can be interpreted as particle. Rather, it consists of the
continuum spectrum with a mass gap. To see it, we consider the
boundary correlation function without the spinor structure for
simplicity. For the fixed spatial momentum $p_1$,
the correlation function can be evaluated as
\begin{eqnarray*}
D(n_2;p_1)
&=& \int\frac{d p_2}{(2\pi)} e^{i p_2 n_2} \frac{1}{2 \lambda_-} \\
&=& \int\frac{d p_2}{(2\pi)}\frac{d \omega}{(2\pi)}
 e^{i p_2 n_2} \frac{1}{\omega^2+ \lambda_-^2} \\
&=& \int \frac{d \omega}{(2\pi)}
\frac{1}{(2-\cos p_1 - m_0)}
\frac{ e^{-M(\omega,p_1) |n_2|}}{2 \sinh M(\omega,p_1)} ,
\end{eqnarray*}
where
\begin{equation}
 \cosh M(\omega,p_1) =
1 +
\frac{ \omega^2 +\sin^2 p_1+ (1-\cos p_1 - m_0)^2}{2 (2-\cos p_1 - m_0)} .
\end{equation}
The minimum of $M(\omega,p_1)$ can be identified as the mass gap.
For $m_0 = 0.5$, it appears at $\omega=0$ and $p_1=0$.
In this case,  the mass gap $M_B$ is given by
\begin{equation}
\cosh M_B = 1 + \frac{ m_0^2}{2 (1 - m_0)} .
\end{equation}
Since the mass gap is of order of the cutoff,
no light physical particle emerges
in the boundary correlation functions on the critical point.
They have very short-distance nature.
\footnote{In fact, the boundary correlation function can be derived
from the correlation function of the three-dimensional massive
fermion by the reduction to the two-dimensional space-time.
This is why the continuum spectrum with mass gap emerges. }

\subsection{Boundary correlation functions off criticality}

Once we know that the boundary correlation functions have
short distance nature, the quantum correction to such quantities
can be evaluated by the perturbation theory in $\lambda$ rather
reliably by virtue of the asymptotic freedom.

In the perturbation theory in $\lambda$, $S_+^v[g](n,m)$ can
be expanded as follows:
\begin{eqnarray}
S_-^v[g](n,m){}_s^o
&=&
S^v_-( n-m ) \left( \bbone{}_s^o \right)
-
\sum_r S^v_-( n-r ) \left( (-i\lambda) \pi_r{}_s^o \right) S^v_-( r-m )
\nonumber\\
&&
-\sum_r S^v_-( n-r )
\left( \frac{(-i\lambda)^2}{2!}\pi_r^2{}_s^o \right) S^v_-( r-m )
\nonumber\\
&&
+ \sum_{r,l}
S^v_-( n-r ) \left( (-i\lambda) \pi_r{}_s^u \right)
S^v_-( r-l ) \left( (-i\lambda) \pi_l{}_u^o \right)
S^v_-( l-m )
\nonumber\\
&&
+{\cal O}(\lambda^3) .
\end{eqnarray}
Then we obtain at the one-loop order
\begin{eqnarray}
  \langle \phi_n{}_i \phi^\dagger_m{}^i  \rangle_{-r}
&=&
\frac{1}{2} \delta_{nm} \delta_i^i - S^v_-( n-m ) \delta_i^i
\nonumber\\
&-&
\lambda^2  \sum_{r}
S^v_-( n-r )
\left[
\langle \pi_r{}_i^o \pi_m{}_o^i \rangle^\prime S^v_-( r-m )
\right]
\nonumber\\
&+&
\lambda^2  \sum_{r,l}
S^v_-( n-r )
\left[
\langle \pi_r{}_i^o \pi_l{}_o^i \rangle^\prime S^v_-( r-l )
\right] S^v_-( l-m ) +{\cal O}(\lambda^4) ,
\nonumber\\
\label{eq:boundary-correlation-function-colored}
&&\\
  \langle \varphi_n{}_i \varphi^\dagger_m{}^i  \rangle_{-r}
&=&
\frac{1}{2} \delta_{nm} \delta_i^i - S^v_-( n-m ) \delta_i^i
\nonumber\\
&-&
\lambda^2  \sum_{r}
\left[
\langle \pi_n{}_i^o \pi_r{}_o^i \rangle^\prime S^v_-( n-r )
\right]
S^v_-( r-m )
\nonumber\\
&+&
\lambda^2  \sum_{r,l}
S^v_-( n-r )
\left[
\langle \pi_r{}_i^o \pi_l{}_o^i \rangle^\prime S^v_-( r-l )
\right] S^v_-( l-m ) +{\cal O}(\lambda^4) ,
\nonumber\\
\label{eq:boundary-correlation-function-noncolored}
&&\\
\langle \pi_r{}_i^o \pi_l{}_o^i \rangle^\prime
&=& \delta_i^i \int \frac{d^2 p}{(2\pi)^2}
\frac{ e^{i p(r-l) } -1 }
{\sum_\mu 4 \sin^2 \frac{p_\mu}{2}} .
\end{eqnarray}
Note that the infrared divergences associated to the correlation
function of the gauge freedom
\begin{equation}
\langle \pi_r{}_i^o \pi_l{}_o^i \rangle
= \delta_i^i \int \frac{d^2 p}{(2\pi)^2}
\frac{ e^{i p(r-l) } }
{\sum_\mu 4 \sin^2 \frac{p_\mu}{2}} ,
\end{equation}
cancel among the second and third terms in the r.h.s. of
Eqs.~(\ref{eq:boundary-correlation-function-colored}) and
(\ref{eq:boundary-correlation-function-noncolored}).
To show this fact explicitly, we have used
$\langle \pi_r{}_i^o \pi_l{}_o^i \rangle^\prime $ insead.\footnote{
It is not hard to see that the infrared divergence remains
in the correlation function of the type of
Eq.~(\ref{eq:boundary-correlation-function-gauge-mixed}). }

Since $S_-^v$ has the short correlation length of the order of the
lattice spacing as we have shown,
its combolutions with the correlation function of the gauge freedom
also have the short correlation lengths.
There is no symmetry against the mass gap.
This result shows that even inside of the symmetric phase
off the critical point, no light particle emerges
in the boundary correlation functions.

\section{Decoupling of gauge degree of freedom}
\label{sc:decoupling-of-gauge-freedom}
\reseteqnum

The mass of the gauge freedom, which we denote by $M_\sigma$,
is adjustable by $\lambda$.  Near the critical point,
the scale $M_\sigma$ is very small compared to the lattice cutoff.
Here, we found the massless {\it Weyl} fermion in the overlap
correlation function. On the other hand, the spectrum in
the boundary correlation function has the large mass gap $M_B$
of the order of the lattice cutoff. This mass scale is set by
the constant $m_0$ and not adjustable.
The mass gap suffers from the quantum correction due to the
fluctuation of the gauge degree of freedom, but survives it.
Therefore we obtain the following relations of several scales for
small $\lambda$:
\begin{equation}
0 \, (= \Lambda[\beta])  \simeq M_\sigma[\lambda]
\, \, << \, \, M_B[m_0;\lambda] \simeq \frac{1}{a}
      \qquad (\lambda^2 \sim 0) .
\end{equation}

How far can we make the mass $M_\sigma$ large?
As $\lambda$ becomes large, $M_\sigma$ becomes large.
At the same time, the fluctuation of the gauge degree of freedom
increases. Then, the quantum correction to the mass gap of the boundary
correlation function can become large.
However, we do not find any symmetrical reason why it could become
vanishingly small compared to the lattice cutoff.
Then, it seems quite reasonable
to assume that this short distance nature remains as $\lambda$ becomes
large. This means that we can make $M_\sigma$ large and comparable
to $M_B$ from below.
\begin{equation}
0 \, (= \Lambda[\beta])  \, \, << \, \,
M_\sigma[\lambda]  \nearrow
M_B[\lambda;m_0] \simeq \frac{1}{a}  \qquad (\lambda^2 \nearrow \infty ) .
\end{equation}
Then all the unphysical and undesired degrees of freedom
could be heavy compared to the physical scale $\Lambda[\beta]$.

The important assumption we have made here is that
no phase transition occurs between the week coupling region
and the strong coupling region of $\lambda$.
In the chiral $SU(2)$ nonlinear sigma model, it is known to be true.
The disordered nature at strong coupling holds
true in the week coupling regime.
By the spin wave approximation, we see that
the correlation function does not show the long range order and
the local order parameter does not emerges.
Rather, due to the asymptotic freedom, the mass is
developed dynamically.

The week coupling dynamics of the pure gauge model in consideration,
has also disordered nature, as we have shown.
As $\lambda$ increases, the disordered nature of chiral field is
enhanced.  But in our pure gauge model, this means that the imaginary
null Wess-Zumino action can become large and can fluctuate strongly.
Our assumption here
is that {\it as far as an anomaly-free chiral gauge theory is concerned,
there is no phase transition which divides up the weak
and the strong coupling regions of the asymptotically free
self-coupling of the gauge freedom}.

An important and well-known counter example occurs
if the gauge anomaly does not cancel and the Wess-Zumino
term appears in the complex action. The Wess-Zumino term causes
the IR fixed point in the beta function of $\lambda $\cite{witt}.
It has been shown to be also true in the vacuum overlap
formulation\cite{wess-zumino-term-overlap}.
At one-loop order, the beta function is given by
\begin{equation}
- a \frac{d \lambda^2}{d a }
= - \left( 1- A \frac{\lambda^2}{8\pi} \right) \frac{\lambda^4}{4\pi} .
\end{equation}
The fixed point theory is equivalent to the free massless fermion
with the chiral $SU(2)$ symmetry\cite{witt}.  This simply means
the failure of the decoupling of the gauge degree of freedom.
This freedom appears as extra light particles at the fixed point.
As in this example, the nature of the pure gauge dynamics
could distinguish the anomaly-free chiral gauge theory from
anomalous ones.

\section{Effect of boundary Yukawa coupling}
\label{sc:boundary-Yukawa-coupling}
\reseteqnum

In this section, we try to understand the effect of the Yukawa
coupling at the waveguide boundary\cite{waveguide-analysis-golterman}
in the context of the vacuum overlap. We will find that
as far as $y$ is kept nonzero finite, the mass gap of a boundary
correlation function remains finite.

\subsection{Boundary Yukawa coupling}
The Yukawa coupling introduced at the waveguide
boundary\cite{waveguide-analysis-golterman}
corresponds to the insertion of the following
operator\cite{boundary-yukawa-coupling-operator}
in the overlaps to define the complex phase:
\begin{eqnarray}
 \hat Y
&=&
\exp \left(
\sum_{n,i} \ln y \, ( \hat a_{n 1}^\dagger{}^i \hat a_{n 1}{}_i
                    - \hat a_{n 2}^\dagger{}^i \hat a_{n 2}{}_i )
     \right)
\nonumber\\
&=&
 \prod_{n i}
\left(   \hat a_{n 1}{}_i \hat a_{n 1}^\dagger{}^i
     + y \hat a_{n 1}^\dagger{}^i \hat a_{n 1}{}_i  \right)
\left( \hat a_{n 2}{}_i \hat a_{n 2}^\dagger{}^i
     + \frac{1}{y} \hat a_{n 2}^\dagger{}^i \hat a_{n 2}{}_i \right) ,
\end{eqnarray}
and
\begin{eqnarray}
    Z&= &
\int [dg]
\exp \left\{
K \sum_{n\mu} {\rm Tr}( g_n g_{n+\mu}^\dagger  + g_{n+\mu}g_n^\dagger )
     \right\}
      \prod_{rep.} \left(
   \frac{\lvacp \vert \hat Y \hat G \rvacp}
  {|\lvacp \vert \hat Y \hat G \rvacp|}
   \lvacp \rvacm
   \frac{\lvacm \vert \hat G^\dagger \hat Y \rvacm}
        {|\lvacm \vert \hat G^\dagger \hat Y \rvacm|}
              \right)
\nonumber\\
&\equiv& \int d\mu[g;K,Y] .
\end{eqnarray}
We denote by $Y$ the matrix in the spinor space given by
\begin{equation}
    Y = \left(
       \begin{array}{cc} y &  0 \\ 0 & \frac{1}{y} \end{array}
        \right)
      = y P_R + \frac{1}{y} P_L .
\end{equation}

\subsection{Null Wess-Zumino action with Yukawa coupling}

We first clarify the effect of the boundary Yukawa coupling
in the null Wess-Zumino action. The boundary Yukawa coupling
comes in the explicit formula of
the null Wess-Zumino action as follows:
\begin{eqnarray}
i\Delta\Gamma_{WZ}[g;Y]_r
&=&
i {\rm Im} {\rm Tr} \, {\rm Ln} \left[
\sum_{m}
v^\dagger_-(p) e^{-i p m}
\left( g_m^\dagger{}_i^j \right)  e^{i q m } Y v_-(q)
\right]
\nonumber\\
&+&
i {\rm Im} {\rm Tr} \, {\rm Ln} \left[
\sum_{m}
v^\dagger_+(p) Y e^{-i p m}
\left( g_m{}_i^j \right)  e^{i q m }  v_+(q)
\right] .
\nonumber\\
\end{eqnarray}

In the limits $y=0$ and $y=\infty$,
the gauge freedom decouples and the null Wess-Zumino action vanishes.
In the limit $y=0$, for example, the wavefunctions of
the eigenvectors reduce to only lower components.
\begin{eqnarray}
v_-(p) &\rightarrow&
\frac{1}{y}
\left(  \begin{array}{c} 0 \\ C^\dagger(p) \end{array} \right)
  \frac{1}{\sqrt{2\lambda_-(\lambda_- + m_0-B(p))}} ,
\\
v_+(p)&\rightarrow&
\frac{1}{y}
\left( \begin{array}{c}
       0 \\
       -B(p)-m_0-\lambda_- \end{array} \right)
  \frac{1}{\sqrt{2\lambda_+(\lambda_+ + m_0 + B(p))}}.
\end{eqnarray}
Then the wavefunctions factorize in the formula of the null
Wess-Zumino action as follows:
\begin{eqnarray}
&&
{\rm Tr} \, {\rm Ln} \left[
\sum_{m}
v^\dagger_-(p) e^{-i p m}
\left( g_m^\dagger{}_i^j \right)  e^{i q m } Y v_-(q) \right]
\nonumber\\
&&
=
{\rm Tr} \, {\rm Ln} \left[
\sum_{m}
e^{-i p m}
\left( g_m^\dagger{}_i^j \right)  e^{i q m } \right]
\nonumber\\
&&
+{\rm Tr} \, {\rm Ln} \left[ v^\dagger_-(p){}_2
(2\pi)^2 \delta^2(p-q)
\right]
+{\rm Tr} \, {\rm Ln} \left[ (2\pi)^2 \delta^2(p-q)
\frac{1}{y} v_-(q){}_2 \right] .
\nonumber\\
\end{eqnarray}
The first term in the r.h.s. vanishes because $g_n{}_i^j$ are the special
unitary matrices. The second and third term in the r.h.s. combines to give
the real value. In this way, the gauge freedom decouples from the
fermion determinant and the null Wess-Zumino action vanishes.
In the limit $y=\infty$, the wavefunctions of the eigenvectors
reduce to only upper components and the gauge freedom decouples
in the similar manner.

Perturbative evaluation of the null Wess-Zumino action performs
just like the case without the boundary Yukawa coupling.
The term of the order ${\cal O}(\pi^3)$ is evaluated as
\begin{eqnarray}
&&
i \Delta\Gamma_{WZ}[\pi;Y]_{r-}
\nonumber\\
&&
= i \frac{1}{3} \lambda^3
\int \frac{d^2 p_1}{(2\pi)^2}
\frac{d^2 p_2}{(2\pi)^2} \frac{d^2 p_3}{(2\pi)^2}
(2\pi)^2 \delta^2(p_1+p_2+p_3)
i \epsilon^{abc} \left( \pi^a(p_1) \pi^b(p_2) \pi^c(p_3) \right)
\times
\nonumber\\
&& \qquad \qquad
A_r
\int \frac{d^2 q}{(2\pi)^2}
{\rm Tr} \left[ S^v_-(q+p_1;Y)S^v_-(q;Y)S^v_-(q-p_2;Y) \right] + \cdots .
\end{eqnarray}
where
\begin{equation}
  S^v_-(p;Y) \equiv \frac{1}{v^\dagger_-(p)Y v_-(p)}
   Y v_-(p) v^\dagger_-(p) .
\end{equation}
Therefore, by the anomaly cancellation, the induced null Wess-Zumino
action also vanishes completely up to ${\cal O}(\pi{}^3)$.\footnote{
In the wave-guide formulation, the induced action for the gauge freedom
from the fermion determinants has a real part and depends on the
boundary Yukawa coupling. It affects the phase structure of the pure
gauge model. No such correction occurs in the vacuum overlap
formulation and this simplifies the analysis of the phase structure.}

The coefficient of the Wess-Zumino term, $c_{WZ}[Y]_-$, now seems to
depend on $Y$. It is evaluated again using
technique of \cite{chern-simons-current} as follows:
\begin{eqnarray}
&&
c_{WZ}[Y]_-
\nonumber\\
&&=
\int \frac{d^2 q}{(2\pi)^2}
\frac{1}{2!} \epsilon_{\mu\nu}
{\rm Tr} \left[
\partial_\mu S^v_-(q;Y) S^v_-(q;Y) \partial_\nu S^v_-(q;Y)
\right]
\nonumber\\
&&= (-i) \frac{1}{2}
\int \frac{d^2 q}{(2\pi)^2} \int_{-\infty}^\infty \frac{d \omega}{(2\pi)}
\nonumber\\
&& \quad
\frac{1}{2!} \epsilon_{\mu\nu}
{\rm Tr} \left\{ S_3(q,\omega;Y) \partial_\mu S_3^{-1}(q,\omega;Y) \times
\right. \nonumber\\
&&\left. \qquad\qquad\qquad
                 S_3(q,\omega;Y) \partial_\nu S_3^{-1}(q,\omega;Y)
                 S_3(q,\omega;Y) \partial_\omega S_3^{-1}(q,\omega;Y)
         \right\}
\nonumber\\
&&=  (-i) \frac{1}{2}  \frac{1}{4\pi} \sum_{k=0}^{2} (-1)^k
      \left(
      \begin{array}{c} 2 \\ k \end{array} \right)
      \frac{ m_k } { \vert m_k \vert }
\qquad m_k= \left\{
  \begin{array}{c} -y m_0  \quad (k=0) \\
                   \frac{1}{y} (2k-m_0) \quad (k\not=0) \end{array} \right.
\nonumber\\
&&=(-i) \frac{1}{4\pi}  .
\end{eqnarray}
where
\begin{eqnarray}
S_3^{-1}(p,\omega; Y)
&\equiv &
i \gamma_5 \omega + i \gamma_\mu C_\mu(p) +
\frac{1}{2}\left[ y(B-m_0-\lambda_-)
                  +\frac{1}{y} (B-m_0+\lambda_-) \right] .
\nonumber\\
\end{eqnarray}
For the positive mass case, we obtain by the similar calculation
\begin{equation}
c_{WZ}[Y]_+ =0 .
\end{equation}
Thus, as far as $y$ is kept finite, it does not depend on $y$ and
reproduces the correct coefficient.
It is only when $y=0$ or $y=\infty$,
the gauge freedom decouples and the Wess-Zumino term vanishes.
This means that the coefficient of the Wess-Zumino term
shows nonanalytic behavior at $y=0$ and $y=\infty$.

\subsection{Boundary correlation function with Yukawa coupling}

With the boundary Yukawa coupling, there are four possible definitions
of the invariant boundary correlation functions:
\begin{eqnarray}
\langle \phi_n{}_i \phi^\dagger_m{}^i  \rangle_{-r}^{IN}
&\equiv&\frac{1}{Z} \int d\mu[g;K,Y]
\frac{\lvacm \vert \hat G^\dagger
  \left\{ \hat a_n{}_i \hat a^\dagger_m{}^i
         -\frac{1}{2} \delta_{nm} \delta_i^i \right\}
      \hat Y \rvacm_r}
     {\lvacm \vert \hat G^\dagger \hat Y \rvacm_r } ,
\nonumber\\
&\\
\langle \varphi_n{}_i \varphi^\dagger_m{}^i  \rangle_{-r}^{IN}
&\equiv&\frac{1}{Z} \int d\mu[g;K,Y] \,
\frac{\lvacm \vert
  \left\{ \hat a_n{}_i \hat a^\dagger_m{}^i
         -\frac{1}{2} \delta_{nm} \delta_i^i \right\}
      \hat G^\dagger \hat Y \rvacm_r}
     {\lvacm \vert \hat Y \hat G^\dagger \rvacm_r } ,
\nonumber\\
&\\
\langle \phi_n{}_i \phi^\dagger_m{}^i  \rangle_{-r}^{OUT}
&\equiv&\frac{1}{Z} \int d\mu[g;K,Y]
\frac{\lvacm \vert \hat Y  \hat G^\dagger
  \left\{ \hat a_n{}_i \hat a^\dagger_m{}^i
         -\frac{1}{2} \delta_{nm} \delta_i^i \right\}
     \rvacm_r}
     {\lvacm \vert \hat G^\dagger \hat Y \rvacm_r } ,
\nonumber\\
&\\
\langle \varphi_n{}_i \varphi^\dagger_m{}^i \rangle_{-r}^{OUT}
&\equiv&\frac{1}{Z} \int d\mu[g;K,Y] \,
\frac{\lvacm \vert \hat Y
  \left\{ \hat a_n{}_i \hat a^\dagger_m{}^i
         -\frac{1}{2} \delta_{nm} \delta_i^i \right\}
      \hat G^\dagger \rvacm_r}
     {\lvacm \vert \hat Y \hat G^\dagger \rvacm_r } .
\nonumber\\
\end{eqnarray}
$IN$ and $OUT$ stand for the {\it inside} and {\it outside} of
the waveguide, respectively. In the limit $y=0$, the operator $\hat Y$
plays the role of the projection operator, by which an open boundary
condition can be implemented. In this case, the place to put the
operator $\hat Y$ matters. If it is placed at the most right side of
all operators in-between the overlap, the correlation function is
for the fermion inside the waveguide.
If it is placed at the most left side,
the correlation function is for the fermion outside the waveguide.
We keep these names for a generic $y$, although there is no
such clear separation.
For $y=1$, $IN$ and $OUT$
degenerate. For each case of $IN$ and $OUT$, there are two possible
definitions according to their transformation properties under
the chiral $SU(2)$.

Their explicit expressions are evaluated as
\begin{eqnarray}
\label{eq:boundary-correlation-function-gauge-yukawa-IN-colored}
\langle \phi_n{}_i \phi^\dagger_m{}^i  \rangle_{-r}^{IN}
&=&\frac{1}{Z} \int d\mu[g;K,Y]
\left[ \frac{1}{2} \delta_{nm} \delta_i^i
- S_-^v[g;Y](n;m){}_i^o \left( g^\dagger_m{}_o^i \right)
\right] ,
\nonumber\\
&\\
\label{eq:boundary-correlation-function-gauge-yukawa-IN-noncolored}
\langle \varphi_n{}_i \varphi^\dagger_m{}^i  \rangle_{-r}^{IN}
&=&\frac{1}{Z} \int d\mu[g;K,Y]
\left[ \frac{1}{2} \delta_{nm} \delta_i^i
- \left( g^\dagger_m{}_i^o \right) S_-^v[g;Y](n;m){}_o^i
\right] ,
\nonumber\\
\end{eqnarray}
\begin{eqnarray}
\label{eq:IN-to-OUT}
\langle \phi_n{}_i \phi^\dagger_m{}^i  \rangle_{-r}^{OUT}
&=& Y^{-1} \langle \phi_n{}_i \phi^\dagger_m{}^i  \rangle_{-r}^{IN} Y ,
\\
\langle \varphi_n{}_i \varphi^\dagger_m{}^i  \rangle_{-r}^{OUT}
&=&Y^{-1} \langle \varphi_n{}_i \varphi^\dagger_m{}^i
\rangle_{-r}^{IN} Y ,
\end{eqnarray}
where
\begin{eqnarray}
\label{eq:Chiral-projection-gauge-yukawa}
S_-^v[g](n,m;Y){}_i^j
&\equiv&
\int \frac{d^2 p}{(2\pi)^2} \frac{d^2 q}{(2\pi)^2} \times
\nonumber\\
&&
e^{i p n } Y v_-(p)
\left[ v^\dagger_-(q)
e^{-i q r} \left( g_r^\dagger{}_j^i \right) e^{i p r }
      Y v_-(p) \right]^{-1}_{(p,i)(q,j)}
v^\dagger_-(q) e^{-i q m} .
\nonumber\\
\end{eqnarray}

\subsubsection{Boundary correlation functions in
               the limits $y=0$ and $y=\infty$}

We first examine the boundary correlation functions in the limits
$y=0$ and $y=\infty$. In the case $y << 1$, it is useful to rewrite
Eq.~(\ref{eq:Chiral-projection-gauge-yukawa}) in the following form:
\begin{eqnarray}
\label{eq:Chiral-projection-gauge-yukawa-ready-to-expansion-in-y}
&&S_-^v[g](n,m;Y){}_i^j
\nonumber\\
&&=
\int \frac{d^2 p}{(2\pi)^2} \frac{d^2 q}{(2\pi)^2}
\times
\nonumber\\
&&
e^{i p n }
\left( \begin{array}{c}
          y^2 \left( B(p)-m_0-\lambda_-(p) \right) \\
          C^\dagger(p)
       \end{array} \right)
\left( \begin{array}{cc}
    B(q)-m_0-\lambda_-(q) & C(q)
       \end{array} \right) e^{-i q m}
\times
\nonumber\\
&&
\left\{
(2\pi)^2 \delta(p_s-p) \delta_s^i
\phantom{\frac{1}{C^\dagger(p_s)}}
\right.
\nonumber\\
&& \quad
+ y^2
\frac{1}{C^\dagger(p_s)} \,
   e^{-i p_s r} \left( g_r{}_s^t \right) e^{i p_t r }
\frac{\left(B(p_t)-m_0-\lambda_-(p_t) \right)}
     {C(p_t)} \times
\nonumber\\
&& \left.
\quad\qquad\qquad\phantom{\frac{1}{C^\dagger(p_s)}}
   e^{-i p_t l} \left( g_{l}^\dagger{}_t^i \right) e^{i q l }
    \left(B(p)-m_0-\lambda_-(p) \right) \,
\right\}^{-1}_{(p,i)(p_s,s)}
\times
\nonumber\\
&& \quad
\frac{1}{C^\dagger(p_s)}
e^{-i p_s r} \left( g_r{}_s^j \right) e^{i q r }
\frac{1}{C(q)} .
\end{eqnarray}
{}From this expression, we can see what happens in the limit $y=0$.
The boundary correlation functions turn out to be
\begin{eqnarray}
\langle \varphi_n{}_i \varphi^\dagger_m{}^i  \rangle_{-r}^{IN}
&=&
\frac{1}{2} \delta_{nm} \delta_i^i
- \delta_i^i
\int \frac{d^2 p}{(2\pi)^2} e^{i p (n-m) }
\left( \begin{array}{cc}
          0 & 0 \\
-\frac{C^\dagger(p)}{B(p)-m_0+\lambda_-(p)} & 1
       \end{array} \right) ,
\nonumber\\
\\
\langle \phi_n{}_i \phi^\dagger_m{}^i  \rangle_{-r}^{IN}
&=&\langle g_n{}_i^s  g^\dagger_m{}_t^i  \rangle
   \langle \varphi_n{}_s \varphi^\dagger_m{}^t  \rangle_{-r}^{IN} ,
\nonumber\\
\\
\langle \phi_n{}_i \phi^\dagger_m{}^i  \rangle_{-r}^{OUT}
&=&
\frac{1}{2} \delta_{nm} \delta_i^i
- \delta_i^i
\int \frac{d^2 p}{(2\pi)^2} e^{i p (n-m) }
\left( \begin{array}{cc}
          0 & - \frac{C(p)}{B(p)-m_0+\lambda_-(p)} \\
          0 & 1
       \end{array} \right) ,
\nonumber\\
\\
\langle \varphi_n{}_i \varphi^\dagger_m{}^i  \rangle_{-r}^{OUT}
&=&
\langle g^\dagger_n{}_i^s  g_m{}_t^i \rangle
\langle \phi_n{}_s \phi^\dagger_m{}^t  \rangle_{-r}^{OUT} .
\nonumber\\
\end{eqnarray}
The gauge freedom decouples from
$\langle\varphi_n{}_i \varphi^\dagger_m{}^i\rangle_{-r}^{IN}$
and  $\langle\phi_n{}_i \phi^\dagger_m{}^i\rangle_{-r}^{OUT}$.
$\langle\phi_n{}_i \phi^\dagger_m{}^i\rangle_{-r}^{IN}$ and
$\langle\varphi_n{}_i \varphi^\dagger_m{}^i\rangle_{-r}^{OUT}$
are constructed from
$\langle\varphi_n{}_i \varphi^\dagger_m{}^i\rangle_{-r}^{IN}$ and
$\langle\phi_n{}_i \phi^\dagger_m{}^i\rangle_{-r}^{OUT}$,
respectively by the combolutions with the correlation functions
of the gauge freedom.

In $\langle\varphi_n{}_i \varphi^\dagger_m{}^i\rangle_{-r}^{IN}$,
only the left-handed component can propagate.
The massless pole appears at $p_\mu=(0,0)$ in the factor
\begin{equation}
  \frac{1}{B(p)-m_0 + \lambda_-} .
\end{equation}
Therefore a single left-handed {\it Weyl} fermion emerges.
For the vanishing momenta for the species doublers, the correlation
function vanishes.
Similarly, in $\langle\phi_n{}_i \phi^\dagger_m{}^i\rangle_{-r}^{OUT}$,
only the right-handed component can propagate and
a single right-handed {\it Weyl} fermion emerges.
In $\langle\varphi_n{}_i \varphi^\dagger_m{}^i\rangle_{-r}^{IN}$
and  $\langle\phi_n{}_i \phi^\dagger_m{}^i\rangle_{-r}^{OUT}$,
such massless particle can not appear because of the combolution,
except at the critical point $\lambda=0$.

In the case $y >> 1$, it is useful to rewrite
Eq.~(\ref{eq:Chiral-projection-gauge-yukawa}) in the following form:
\begin{eqnarray}
\label{eq:Chiral-projection-gauge-yukawa-ready-to-expansion-in-1/y}
&&S_-^v[g](n,m;Y){}_i^j
\nonumber\\
&&=
\int \frac{d^2 p}{(2\pi)^2} \frac{d^2 q}{(2\pi)^2}
\times
\nonumber\\
&&
e^{i p n }
\left( \begin{array}{c}
          B(p)-m_0-\lambda_-(p) \\
          \frac{1}{y^2} C^\dagger(p)
       \end{array} \right)
\left( \begin{array}{cc}
    B(q)-m_0-\lambda_-(q) & C(q)
       \end{array} \right) e^{-i q m}
\times
\nonumber\\
&&
\left\{
(2\pi)^2 \delta(p_s-p) \delta_s^i
\phantom{\frac{1}{C^\dagger(p_s)}}
\right.
\nonumber\\
&&
+ \frac{1}{y^2}
\frac{1}{\left(B(p_s)-m_0-\lambda_-(p_s) \right)}
   e^{-i p_s r} \left( g_r{}_s^t \right) e^{i p_t r } \times
\nonumber\\
&& \left.
\qquad
\frac{C(p_t)}
     {\left(B(p_t)-m_0-\lambda_-(p_t) \right)}
   e^{-i p_t l} \left( g_{l}^\dagger{}_t^i \right) e^{i p l }
    C^\dagger(p)
\right\}^{-1}_{(p,i)(p_s,s)}
\times
\nonumber\\
&&
\frac{1}{\left(B(p_s)-m_0-\lambda_-(p_s) \right)}
e^{-i p_s r} \left( g_r{}_j^s \right) e^{i q r }
\frac{1}{\left(B(q)-m_0-\lambda_-(q) \right)} .
\end{eqnarray}
With the help of this expression, we obtain in
the limit $y=\infty$:
\begin{eqnarray}
\langle \varphi_n{}_i \varphi^\dagger_m{}^i  \rangle_{-r}^{IN}
&=&
\frac{1}{2} \delta_{nm} \delta_i^i
- \delta_i^i
\int \frac{d^2 p}{(2\pi)^2} e^{i p (n-m) }
\left( \begin{array}{cc}
          1 & \frac{C(p)}{B(p)-m_0-\lambda_-(p)} \\
          0 & 0
       \end{array} \right) ,
\nonumber\\
\\
\langle \phi_n{}_i \phi^\dagger_m{}^i  \rangle_{-r}^{IN}
&=&
\langle g_n{}_i^s  g^\dagger_m{}_t^i  \rangle
\langle \varphi_n{}_s \varphi^\dagger_m{}^t  \rangle_{-r}^{IN} ,
\nonumber\\
\\
\langle \phi_n{}_i \phi^\dagger_m{}^i  \rangle_{-r}^{OUT}
&=&
\frac{1}{2} \delta_{nm} \delta_i^i
- \delta_i^i
\int \frac{d^2 p}{(2\pi)^2} e^{i p (n-m) }
\left( \begin{array}{cc}
          1 & 0 \\
          \frac{C^\dagger(p)}{B(p)-m_0-\lambda_-(p)} & 0
       \end{array} \right) ,
\nonumber\\
\\
\langle \varphi_n{}_i \varphi^\dagger_m{}^i  \rangle_{-r}^{OUT}
&=&
\langle g^\dagger_n{}_i^s  g_m{}_t^i  \rangle
\langle \phi_n{}_s \phi^\dagger_m{}^t  \rangle_{-r}^{OUT} .
\nonumber\\
\end{eqnarray}
The decoupling of the gauge freedom again occurs to
$\langle\varphi_n{}_i \varphi^\dagger_m{}^i\rangle_{-r}^{IN}$
and  $\langle\phi_n{}_i \phi^\dagger_m{}^i\rangle_{-r}^{OUT}$.
$\langle\phi_n{}_i \phi^\dagger_m{}^i\rangle_{-r}^{IN}$ and
$\langle\varphi_n{}_i \varphi^\dagger_m{}^i\rangle_{-r}^{OUT}$
are again given by the combolutions.
In this case, the massless spectrum drastically changes.
In $\langle\varphi_n{}_i \varphi^\dagger_m{}^i\rangle_{-r}^{IN}$,
only the right-handed component can propagate.
The massless pole appears at $p_\mu=(0,\pi)$,$(\pi,0)$ and
$(\pi,\pi)$ in the factor
\begin{equation}
  \frac{1}{B(p)-m_0 - \lambda_-} .
\end{equation}
At $p_\mu=(0,0)$, the correlation function vanishes.
Therefore three right-handed {\it Weyl} fermion emerge.
Similarly,
in $\langle\phi_n{}_i \phi^\dagger_m{}^i\rangle_{-r}^{OUT}$,
only the left-handed component can propagate and three left-handed
{\it Weyl} fermions emerge.
All these results of the massless fermions
in the limits $y=0$ and $y=\infty$
is completely consistent with the result
of \cite{waveguide-analysis-golterman}.

\subsubsection{Boundary correlation function on and off the criticality}
Next we examine the boundary correlation functions with a generic
Yukawa coupling in the vicinity of the critical point of the gauge
freedom.

At the first order of the perturbative expansion in $\lambda$,
the IN boundary correlation functions are evaluated as follows.
\begin{eqnarray}
\label{eq:Boundary-correlation-Y-expansion-in-lambda}
  \langle \phi_n{}_i \phi^\dagger_m{}^i  \rangle_{-r}^{IN}
&=&
\frac{1}{2} \delta_{nm} \delta_i^i - S^v_-( n-m;Y ) \delta_i^i
\nonumber\\
&-&
\lambda^2  \sum_{r}
S^v_-( n-r ;Y)
\left[
\langle \pi_r{}_i^o \pi_m{}_o^i \rangle^\prime S^v_-( r-m;Y )
\right]
\nonumber\\
&+&
\lambda^2  \sum_{r,l}
S^v_-( n-r;Y )
\left[
\langle \pi_r{}_i^o \pi_l{}_o^i \rangle^\prime S^v_-( r-l;Y )
\right] S^v_-( l-m;Y ) +{\cal O}(\lambda^4) ,
\nonumber\\
  \langle \varphi_n{}_i \varphi^\dagger_m{}^i  \rangle_{-r}^{IN}
&=&
\frac{1}{2} \delta_{nm} \delta_i^i - S^v_-( n-m;Y ) \delta_i^i
\nonumber\\
&-&
\lambda^2  \sum_{r}
\left[
\langle \pi_n{}_i^o \pi_r{}_o^i \rangle^\prime S^v_-( n-r;Y )
\right]
S^v_-( r-m;Y )
\nonumber\\
&+&
\lambda^2  \sum_{r,l}
S^v_-( n-r;Y )
\left[
\langle \pi_r{}_i^o \pi_l{}_o^i \rangle^\prime S^v_-( r-l;Y )
\right] S^v_-( l-m;Y ) +{\cal O}(\lambda^4) ,
\nonumber\\
\end{eqnarray}
The OUT boundary correlation functions are obtained by the
relation Eq.~(\ref{eq:IN-to-OUT})

At the critical point, they reduce to
\begin{eqnarray}
\langle \phi_n{}_i \phi^\dagger_m{}^i  \rangle_{-r}^{IN}
&=&\langle \varphi_n{}_i \varphi^\dagger_m{}^i  \rangle_{-r}^{IN}
\\
&=&
\delta_i^i \int \frac{d^2 p}{(2\pi)^2} e^{i p (n-m)}
\left[ \frac{1}{2} - S_-^v(p;Y)   \right] ,
\\
\langle \varphi_n{}_i \varphi^\dagger_m{}^i  \rangle_{-r}^{OUT}
&=&\langle \phi_n{}_i \phi^\dagger_m{}^i  \rangle_{-r}^{OUT}
\\
&=&
\delta_i^i \int \frac{d^2 p}{(2\pi)^2} e^{i p (n-m)}
\left[ \frac{1}{2} - Y^{-1} S_-^v(p;Y) Y  \right] ,
\end{eqnarray}
where
\begin{eqnarray}
\label{eq:Chiral-projection-yukawa}
S_-^v(p;Y)
&\equiv& \frac{1}{v^\dagger_-(p)Y v_-(p)}  Y v_-(p) v^\dagger_-(p)
\nonumber\\
&=&
\frac{1}
{\left[y (\lambda_-+m_0-B(p))+\frac{1}{y} (\lambda_--m_0+B(p))\right]}
\times
\nonumber\\
&& \qquad
\left( \begin{array}{cc}
y\left(\lambda_-+m_0-B(p) \right)
&  -y C(p) \\
   -\frac{1}{y} C^\dagger(p)
&
\frac{1}{y} \left( \lambda_- -m_0+B(p)\right)
       \end{array}\right) ,
\nonumber\\
\\
Y^{-1} S_-^v(p;Y) Y
&=&
\frac{1}
{\left[y (\lambda_-+m_0-B(p))+\frac{1}{y} (\lambda_--m_0+B(p))\right]}
\times
\nonumber\\
&& \qquad
\left( \begin{array}{cc}
y\left(\lambda_-+m_0-B(p) \right)
&  -\frac{1}{y} C(p) \\
   -y C^\dagger(p)
&
\frac{1}{y} \left( \lambda_- -m_0+B(p)\right)
       \end{array}\right) .
\nonumber\\
\end{eqnarray}

As the case without the boundary Yukawa coupling,
it seems that these boundary correlation functions
does not show any pole which can be interpreted as particle and
it consists of the continuum spectrum with a mass gap.
However, the structure of the spectrum is more complicated.
As in the previous case,
let us consider the boundary correlation
function without the spinor structure for simplicity.
It can be written as
\begin{eqnarray}
D(n;Y )
&=& \int\frac{d^2 p}{(2\pi)^2} e^{i p n} \frac{1}{2 \lambda_-^\prime} \\
&=& \int\frac{d^2 p}{(2\pi)^2}\frac{d \omega}{(2\pi)}
 e^{i p n} \frac{1}{\omega^2+ \lambda_-^{\prime 2}}  ,
\end{eqnarray}
where
\begin{equation}
  \lambda_-^\prime \equiv \frac{1}{2}
\left[y (\lambda_-+m_0-B(p))+\frac{1}{y} (\lambda_--m_0+B(p))\right] .
\end{equation}
The mass gap of the correlation function is identified as
the minimum of the energy function $E(p_1)$ ($p_1 \in [0,\pi]$)
which solves the equation
\begin{equation}
\label{eq:gap-equation}
\lambda_-^{\prime 2}\left( p_1,p_2= i E(p_1)\right)  = 0.
\end{equation}
In terms of $X = \cosh E(p_1)$, the above equation reads
\begin{eqnarray}
\label{eq:gap-equation-X}
&& \left\{ 1+\sin^2 p_1 +(2-\cos p_1-m_0)^2
-2(2-\cos p_1-m_0) X \right\}
\nonumber\\
&&
+\left( \frac{ \frac{1}{y}-y }{ \frac{1}{y}+y }   \right)^2
  ( 2-\cos p_1 -m_0 - X )^2
\nonumber\\
&&
+ 2 \left( \frac{ \frac{1}{y}-y }{ \frac{1}{y}+y }   \right)
  \lambda_+ ( 2-\cos p_1 -m_0 - X ) = 0 ,
\end{eqnarray}
where
\begin{equation}
\lambda_-
= \sqrt{ 1+\sin^2 p_1 +(2-\cos p_1-m_0)^2-2(2-\cos p_1-m_0) X } .
\end{equation}

\begin{figure}
\epsfxsize=8cm
\centerline{\epsffile{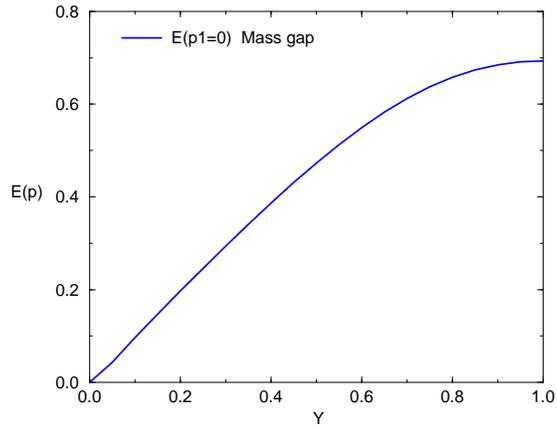}}
\caption{Energy function at $p_1=0$,
         which gives the mass gap for $ y \le 1$, $m_0=0.5$.}
\label{fig:yP}
\end{figure}

For $y \le 1$, a real solution $X$ must satisfy the relation
\begin{equation}
2-\cos p_1 -m_0  \le X \le
\frac{1+\sin^2 p_1 +(2-\cos p_1-m_0)^2}{2(2-\cos p_1-m_0)} .
\end{equation}
Then the allowed region of momentum $p_1$ is restricted to
$p_1 \in [0,p^c]$ where $p^c$ is given by the solution of
\begin{equation}
2-\cos p^c -m_0  =
\frac{1+\sin^2 p^c +(2-\cos p^c-m_0)^2}{2(2-\cos p^c - m_0)} .
\end{equation}
For $m_0=0.5$, $p^c=1.4821$. In this case, the minimum of $E(p_1)$
emerges at $p_1=0$. We show $E(0)$ as a function of $y$ in
figure~\ref{fig:yP}.

\begin{figure}
\epsfxsize=8cm
\centerline{\epsffile{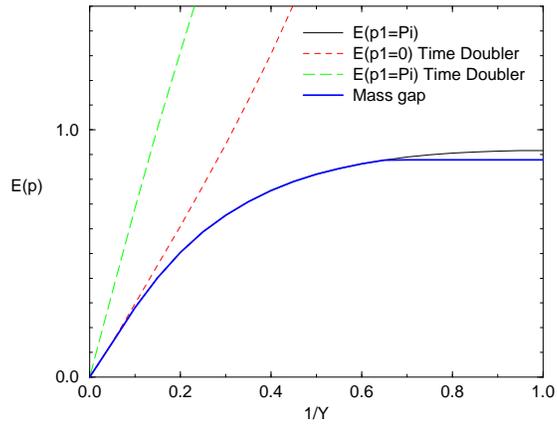}}
\caption{Energy functions of particle at $p_1=\pi$
and time-doublers at $p_1=0,\pi$ for $y \ge 1$, $m_0=0.5$.
The mass gap is also shown. }
\label{fig:YD}
\end{figure}

For $y \ge 1$, one real solution $X$ exists for the momentum in
the range $p_1 \in [p^c,\pi]$.
In figure~\ref{fig:YD}, $E(\pi)$ is shown as a function of $1/y$.
Furthermore, there is a negative real solution
such that $X \le -1$ for each $p_1 \in [0,\pi]$,
which corresponds to the time doubler.
In this case, we define the energy function as
$E(p_1) \equiv \cosh^{-1}(-X)$. For $p_1=0$ and $p_1=\pi$, they are
shown also in figure~\ref{fig:YD}.
The momentum which gives the minimum of $E(p_1)$ shifts as $y$ varies.
For $m_0=0.5$, it shifts from $p^c$ to $\pi$ as $y$ increases. Then
the mass gap merges finally to $E(\pi)$. It is shown also in
figure~\ref{fig:YD}.
Consistently to the previous discussion about the limits
$y=0$ or $y=\infty$, the mass gap of $S_-^v(n-m;Y)$ becomes
small and finally vanishes as the Yukawa coupling becomes
close to those limits.

What actually happens in the limits $y=0$ and $y=\infty$
is worth noting. As clearly seen in
Eq.~(\ref{eq:Chiral-projection-yukawa}),
the wavefunction renormalization constant of one of the chiral
components of the massive correlation functions vanishes
and that component ceases to propagate at the same time the mass
gap vanishes.
In this way, at $y=0$ and $y=\infty$, there emerge
the massless fermions with the chiralities which match vector-likely
with the chirality of the massless {\it Weyl} fermion appearing in
the overlap correlation function.
That vanishing component is naturally understood to
live in the opposite side of the waveguide boundary.
At finite $y$, this is the chiral partner to form massive states.
It has the same transformation property under
the chiral $SU(2)$ and there is no symmetry which forbids the
massive states.

Since the mass gap remains finite for a nonzero finite $y$,
by tuning $\lambda$ close to its critical point, we can
always find the region where the perturbation expansion by $\lambda$,
Eq.~(\ref{eq:Boundary-correlation-Y-expansion-in-lambda}),
is a good approximation to the boundary correlation functions.
The combolutions of $S_-^v(n-m;Y)$ with the correlation function
of the gauge freedom in there also has the finite correlation lengths.
Therefore even inside of the symmetric phase off the critical point,
there emerges no massless particle in the boundary
correlation functions. Again, there is no symmetry against this fact.
\footnote{
Let us comment briefly about our understanding of the numerical
data presented in \cite{waveguide-analysis-golterman}.
Although it is difficult to compare directly, we may get
some qualitative understanding of the data
from our result at the critical point.
In the figure~5 of \cite{waveguide-analysis-golterman},
the inverse propagator of the boundary fermion was shown.
A light states appears there. Compared to
the exact massless state, the light states seems to have
the mass square about 0.11 $(y=0.5,m_0=1.1)$.
According to our result, for $y=0.5$ $(m_0=0.5)$,
it comes out about 0.22 and it is very light.
The dependence of the mass on $y$ was shown in the figure~7 of
\cite{waveguide-analysis-golterman},
which is quite consistent with our result for $ y \le 1$.
In the figure~6 of \cite{waveguide-analysis-golterman},
the dependence of the mass on the coupling constant of the bosonic
field was shown. Although the mass suddenly decreases towards
the critical point,
it is not shown there how the mass becomes vanishingly small
in the symmetric phase.
It seems hard to read the relation $m \propto v$ from this data.
Through the measurement of eigenvalue spectrum and the measurement
of iteration number of CG, no typical qualitative signal
for strong coupling phase was observed.
As we have seen, the Yukawa coupling behaves in a dual way
around $y=1$ and there does not seem to be a region of
strong coupling than $y \simeq 1$ between the ``inside'' and ``outside''
of the (waveguide) boundary. This can be seen from the behavior
of the mass gap of the boundary correlation function.
It is also consistent with the above observed fact.
Therefore the numerical data in \cite{waveguide-analysis-golterman}
does not seem to contradict with our conclusion.
It does not show the clear evidence that the spectrum mass gaps in the
boundary correlation functions vanish in the symmetric phase
for a generic $y$.
}

\subsection{Expansion in $y$ or $1/y$}
Finally, we clarify the structure of the expansions in terms of $y$ and
$1/y$. First we examine at the criticality. As we can see
in Eq.~(\ref{eq:Chiral-projection-yukawa}),
the expansion in terms of $y$  actually means the expansion
in terms of the factor
\begin{equation}
y^2 \frac{\lambda_-+m_0-B(p)}{\lambda_--m_0+B(p)} .
\end{equation}
However, it is never a small number for $p_\mu \simeq (0,0)$,
the expansion breaks down for this region. And in this very region,
the massless pole would appear in the limit $y=0$.
In the case of the expansion in terms of $1/y$, the factor turns out
to be
\begin{equation}
\frac{1}{y^2}\frac{\lambda_--m_0+B(p)}{\lambda_-+m_0-B(p)} .
\end{equation}
It is never a small number for $p_\mu \simeq (\pi,0),(0,\pi),(\pi,\pi)$.
In these regions, the massless poles would appear in the limit $y=\infty$.

For a generic $\lambda$,
as we can see in
Eq.~(\ref{eq:Chiral-projection-gauge-yukawa-ready-to-expansion-in-y}),
the expansion in terms of $y$ means the expansion in terms
of the operator in momentum space:
\begin{eqnarray}
&&
\sum_{r,l,p_t}
y^2
\frac{1}{C^\dagger(p_s)} \,
   e^{-i p_s r} \left( g_r{}_s^t \right) e^{i p_t r }
\frac{\left(B(p_t)-m_0-\lambda_-(p_t) \right)}
     {C(p_t)} \times
\nonumber\\
&&
\quad\qquad\qquad
   e^{-i p_t l} \left( g_{l}^\dagger{}_t^i \right) e^{i q l }
    \left(B(p)-m_0-\lambda_-(p) \right) .
\end{eqnarray}
However, this operator is not bounded. For a generic configuration of
the gauge freedom, it diverges at $p_{s\mu}=(0,0)$.
Similarly,
as we can see in
Eq.~(\ref{eq:Chiral-projection-gauge-yukawa-ready-to-expansion-in-1/y}),
the expansion in terms of $1/y$ means the expansion in terms
of the operator in momentum space:
\begin{eqnarray}
&&
\sum_{r,l,p_t}
  \frac{1}{y^2}
\frac{1}{\left(B(p_s)-m_0-\lambda_-(p_s) \right)}
   e^{-i p_s r} \left( g_r{}_s^t \right) e^{i p_t r } \times
\nonumber\\
&& \quad\qquad
\frac{C(p_t)}
     {\left(B(p_t)-m_0-\lambda_-(p_t) \right)}
   e^{-i p_t l} \left( g_{l}^\dagger{}_t^i \right) e^{i p l }
    C^\dagger(p) .
\end{eqnarray}
This operator is also not bounded. For it diverges at
$p_{s\mu} =(\pi,0)$, $(0,\pi)$ and $(\pi,\pi)$.

Therefore, the expansions in terms of $y$ and $1/y$
break down in the very momentum regions where the massless poles
would appear in the limits $y=0$ and $y=\infty$.
This seems simply because the correlation
function actually has a mass gap and the expansion in terms
of $y$ or $1/y$ means the expansion in terms of the mass gap.
Such expansion is valid only for momenta larger than the mass.
Therefore, we find it hard to claim by these expansions that
the massless poles at $y=0$ and $y=\infty$ persist for nonzero
finite $y$.

\section{Summary and Discussion}
\label{sc:conclusion-discussion}
\reseteqnum

In summary, we have argued that the two requirements for
the pure gauge limit can be fulfilled in the vacuum overlap
formulation of a two-dimensional nonabelian chiral gauge theory.
These requirements are stated as
\begin{enumerate}
\item Mass for the gauge freedom large compared to the physical scale.
\item No physically meaningful light spectrum in boundary
      correlation functions.
\end{enumerate}
In a two-dimensional $SU(2)$ nonablelian chiral gauge theory,
it is naturally expected that the gauge freedom
acquires mass dynamically, because of its two-dimensional and
nonabelian nature.
To make it explicit, we have introduced an asymptotically free
self-coupling for the gauge freedom by hand at first.
We have examined the boundary correlation functions
with the help of the asymptotic freedom and have found that any
other light particle does not emerge there.
Then we have discussed how far the mass of the gauge freedom
can be lifted without affecting the spectrum in the boundary
correlation functions. In this respect, we have pointed out two
facts. There is no symmetry against the spectrum mass gap of
the boundary correlation functions.
The IR fixed point due to the Wess-Zumino term is absent in the
anomaly free theory.

In fact, we have examined several kinds of the boundary
correlation functions. They all have the two-component structure
like the correlation function of the two-dimensional Dirac fermion.
They are all invariant under the chiral $SU(2)$ global symmetry.
There is no symmetry against the mass gap in the spectrum of
these correlation functions.
What we have found is that it is hard to find the massless fermion
in the correlation function from which the gauge degree of freedom
does not decouple by some special reasons.
The gauge freedom does not decouple in general
from the boundary correlation functions. Only when $y=0$ and
$y=\infty$, that is, when some special boundary conditions are imposed,
it happens and the massless {\it Weyl} fermion can emerge there.
Otherwise, they consist of vector-like massive states.
This observation reminds us about
the statement of the Nielsen-Ninomiya theorem extended to the
case of interacting theory by Shamir\cite{NN-Shamir-theorem},
although it might not apply directly to these correlation functions
of nonlocal construction of the vacuum overlap.

In clear contrast,
the decoupling happens in the overlap correlation
functions as shown in \cite{original-overlap}.
There the correct number of free {\it Weyl} fermions emerge.
The global symmetry of the overlap correlation
functions naturally fits the representation of the {\it Weyl}
fermions in the target theory.
The appearance of poles of species doublers are suppressed
by the appearance of zeros at the vanishing momenta for doublers.
This has been pointed out to be a possible way out of the
Nielsen-Ninomiya theorem\cite{NN-theorem, NN-Shamir-theorem}.
Therefore, this decoupling of the gauge freedom can be regarded
as a remarkable property of the definition of the fermion
correlation function in the vacuum overlap formulation.
It is quite parallel to the situation of the
pure gauge limit of the lattice QCD, although the mechanism to
suppress the species doublers are very different.

As for the zeros to suppress the species doublers, however,
the examples are known in which they cause ghost states
which contribute to the vacuum polarization with wrong signature
and lead to the wrong normalization\cite{rebbis,zeros-ghosts}.
In the case of the vacuum overlap formulation,
it has been shown that the perturbative calculation gives
the correct normalization of the vacuum
polarization\cite{saoki-levien,
randjbar-daemi-strathdee,
kikukawa_lattice-vacuum-pol}.
This result is naturally understood from the point of view
of the infinite number of the Pauli-Villars
fields\cite{frolov-slavnov}.
It is still desirable to clarify this point in relation to
the Ward identity\cite{zeros-ghosts}.

The free {\it Weyl} fermions which emerge in the pure
gauge limit do not carry the color indices and does not
transform under the global gauge transformation.
\footnote{
It is important to note that we cannot identify the ``colored'' quark
even in the pure gauge limit of the lattice QCD. This is because
the correlation function of the quark field vanishes according
to the Elitur's theorem\cite{elitur-theorem}.
We may think of the correlation function of the quark field which
is made gauge invariant by the path-ordered product of the link
variables. In the pure gauge limit, it reduces to
the correlation function in discussion. }
They are so-called ``neutral'' fermions in the context of
the Wilson-Yukawa model.
In this context, the coupling of these ``neutral'' fermions
to the vector bosons, which is also gauge singlet, have been discussed
and its triviality was
argued\cite{triviality-wilson-yukawa}.
This is one of the reasons why
the Wilson-Yukawa formulation is not considered to be able to
describe the chiral gauge theory. This argument, however, highly depends
on the dimensions and the dynamical nature of the gauge freedom.
For example, in two-dimensions, the field variable $g_n$ of the
gauge freedom has dimension zero
if it has the self-coupling term which we considered.\footnote{
At the IR fixed point due to the Wess-Zumino term,
the chiral field can be expressed by the free fermion
bilinear operators and has unit mass dimension.
Therefore, the anomaly can also cause difficulty
in this sense.}
In this case, this argument does not lead to the
triviality\cite{two-dimensional-wilson-yukawa}.
In four dimensions, 
if we would consider the higher-derivative coupling for the gauge
freedom as introduced in \cite{renormalizable-gauge-model},
the dimension could differ from unity and
the argument does not seem to lead straightforwardly to the triviality.
Moreover, there is a general question about the physical relevance of
this kind of coupling. Therefore we will leave this issue for
future study and will discuss in more detail elsewhere.

We should also mention about the mean field approximation
in the context of the vacuum overlap formulation.
This approximation replaces the configuration of the
gauge freedom by a constant field,
\begin{equation}
  g_n{}_i^j = v \delta_i^j , \quad g^\dagger_n{}_i^j = v \delta_i^j.
\end{equation}
If we do this replacement rather naively in the vacuum overlap
formulation, then it turns out that the (null) Wess-Zumino action
vanishes and there remains no dependence on $v$ in the fermion sector.
Then we cannot get any information about the anomaly.
This is why we did not discuss with this approximation.

In the two-dimensional $U(1)$ case, however, the Wess-Zumino term
does not exist. Reflecting this fact, the gauge dependence drops out
from the fermion determinant in the vacuum overlap
formulation\cite{original-overlap}. The gauge freedom still
couples to the boundary correlation functions. If we examine them
by the mean field approximation, we find that
the dependence on $v$ also drops out, except for the overall
normalizations.
See
Eqs.~(\ref{eq:boundary-correlation-function-gauge-yukawa-IN-colored}),
(\ref{eq:boundary-correlation-function-gauge-yukawa-IN-noncolored}),
(\ref{eq:Chiral-projection-gauge-yukawa}) and also
Eq.~(\ref{eq:boundary-correlation-function-gauge-mixed}).
\footnote{In the case of $U(1)$, there is no
reason to discard the boundary correlation function
of the type of Eq.~(\ref{eq:boundary-correlation-function-gauge-mixed}).}
This result means that the mass gap in the boundary correlation
function does not depend on the vacuum expectation value. The
proportionality relation between the fermion mass and the vacuum
expectation value, $m = y \, v$, does not hold even in the broken phase.
This might help us to understand what happens in the two-dimensional
$U(1)$ theory.
However, there is still a serious question about this
procedure, because the dependence of the boundary correlation functions
on the field variable of the gauge freedom is nonlocal.

\section*{Acknowledgments}
The author would like to thank H. Neuberger and R. Narayanan
for enlightening discussions. He also would like to thank
S. Aoki and H. So for discussion at early stage of this work.
The author also express his sincere thanks to
High energy physics theory group, Nuclear physics group and the
computer staffs of the department of physics and astronomy
of Rutgers university for their kind hospitality.

\appendix
\vspace{2cm}
\section*{Appendix}

In the following appendixes, we describe in some detail the
evaluations of the formula  discussed in the main text.
In equations, we implicitly mean sum or integration
over the repeated indices or momentum variables.
For the momentum integration, the measure is taken as
$\frac{d^2 p}{(2\pi)^2}$.
We follow closely the second-quantized formulation given
in \cite{randjbar-daemi-strathdee}.

\section{Eigenvectors of free Hamiltonians}
\label{appendix:eigenvectors}
\reseteqnum

In this appendix, we give explicit formula of the eigenvectors
of the free Hamiltonians of the three-dimensional Wilson fermions
with negative and positive masses.
The free Hamiltonians are defined by
  \begin{equation}
    H_{- nm i}^{\phantom{+nm}j}
= \int \frac{d^2 p}{(2\pi)^2} \, e^{i p(n-m) }
\left(\begin{array}{cc}
B(p) - m_0 & i \sigma_\mu \sin p_\mu \\
-i \sigma_\mu^\ast \sin p_\mu  & - B(p) + m_0
      \end{array}\right) \delta_i^j ,
  \end{equation}
  \begin{equation}
    H_{+ nm i}^{\phantom{-nm}j}
= \int \frac{d^2 p}{(2\pi)^2} \, e^{i p(n-m) }
\left(\begin{array}{cc}
B(p) + m_0 & i \sigma_\mu \sin p_\mu \\
-i \sigma_\mu^\ast \sin p_\mu  & - B(p) - m_0
      \end{array}\right) \delta_i^j ,
  \end{equation}
where
\begin{equation}
  C(p) \equiv i \sigma_\mu \sin p_\mu , \quad
  B(p) \equiv \sum_\mu (1-\cos p_\mu ) .
\end{equation}
The eigenvectors are obtained as follows:
  \begin{eqnarray}
u_{-}(p){}_i^k &=&
\delta_i^k
\left( \begin{array}{c}
       C(p) \\
       -B(p)+m_0+\lambda_- \end{array} \right)
  \frac{1}{\sqrt{2\lambda_-(\lambda_- + m_0-B(p))}} ,
\\
v_{-}(p){}_i^k &=&
\delta_i^k
\left( \begin{array}{c}
       B(p)-m_0-\lambda_-  \\
       C^\dagger(p)\end{array} \right)
  \frac{1}{\sqrt{2\lambda_-(\lambda_- + m_0-B(p))}},
  \end{eqnarray}
  \begin{eqnarray}
u_{+}(p){}_i^k &=&
\delta_i^k
\left( \begin{array}{c}
       B(p)+m_0+\lambda_+ \\
       C^\dagger(p) \end{array} \right)
  \frac{1}{\sqrt{2\lambda_+(\lambda_+ + m_0 + B(p))}},
\\
v_{+}(p){}_i^k &=&
\delta_i^k
\left( \begin{array}{c}
       C(p) \\
       -B(p)-m_0-\lambda_+ \end{array} \right)
  \frac{1}{\sqrt{2\lambda_+(\lambda_+ + m_0 + B(p))}} ,
  \end{eqnarray}
where
\begin{equation}
\lambda_- \equiv \sqrt{ C(p)C^\dagger(p) + (B(p)-m_0)^2 } ,
\end{equation}
\begin{equation}
\lambda_+ \equiv \sqrt{ C(p)C^\dagger(p) + (B(p)+m_0)^2 } .
\end{equation}

\section{Vacua of the second-quantized Hamiltonians}
\label{appendix:vacua}
\reseteqnum

In this appendix, we give the vacuum of the second-quantized
Hamiltonian theory of the free three-dimensional Wilson fermion
with negative and positive masses.
The defining creation and annihilation operators
satisfy the following commutation relations.
\begin{eqnarray}
\left\{ \hat a_{ni}, \hat a^\dagger_{m}{}^j \right\}
&=& \delta_{nm} \delta_{ij} ,  \\
\left\{ \hat a_{ni}, \hat a_{mj}\right\}
&=& 0 , \\
\left\{ \hat a^\dagger_{n}{}^i, \hat a^\dagger_{m}{}^j \right\}
&=& 0 .  \\
\end{eqnarray}
The Fock space is constructed on the auxiliary vacuum $\rvac $
defined by
\begin{equation}
    \hat a_n{}_i \rvac = 0 .
\end{equation}

The second quantized Hamiltonian $\hat H_+$ is diagonalized by the
creation and annihilation operators defined as follows:
\begin{eqnarray}
\hat b_+(p){}_k &=& u^\dagger_+(p){}_k^i e^{-i p m} \hat a_m{}_i ,  \\
\hat b_+^\dagger(p){}^k &=& \hat a_n^\dagger{}^i e^{i p n } u_+(p){}_i^k , \\
\hat d_+(p){}^k &=& \hat a_n^\dagger{}^i e^{i p n } v_+(p){}_i^k , \\
\hat d_+^\dagger(p){}_k &=& v^\dagger_+(p){}^i_k e^{-i p m} \hat a_m{}_i ,
\end{eqnarray}
\begin{eqnarray}
\hat a_n^\dagger{}^i  H_{+ nm}{}_i^j \hat a_m{}_j
&=&
\int \frac{d^2 p}{(2\pi)^2} \sum_k \,
\left[
\left( \hat a_n^\dagger{}^i e^{i p n } u_+(p){}_i^k \right)
\lambda_+(p)
\left( u^\dagger_+(p){}^j_k e^{-i p m} \hat a_m{}_j \right)
\right.
\nonumber\\
&& \qquad\qquad\qquad
\left.
-\left( \hat a_n^\dagger{}^i e^{i p n } v_+(p){}_i^k \right)
 \lambda_+(p)
 \left( v^\dagger_+(p){}^j_k e^{-i p m} \hat a_m{}_j \right)
        \right]
\nonumber\\
&=&
\int \frac{d^2 p}{(2\pi)^2} \sum_k \,
        \left[
          \hat b_+^\dagger(p){}^k \lambda_+(p) \hat b_+(p){}_k
       -  \hat d_+(p){}^k \lambda_+(p) \hat d_+^\dagger(p){}_k
        \right] .
\nonumber\\
\end{eqnarray}
Similarly, the second quantized Hamiltonian $\hat H_-$
is diagonalized by the creation and annihilation operators defined
as follows:
\begin{eqnarray}
\hat b_-(p){}_k &=& u^\dagger_-(p){}_k^i e^{-i p m} \hat a_m{}_i ,  \\
\hat b_-^\dagger(p){}^k &=& \hat a_n^\dagger{}^i e^{i p n } u_-(p){}_i^k , \\
\hat d_-(p){}^k &=& \hat a_n^\dagger{}^i e^{i p n } v_-(p){}_i^k , \\
\hat d_-^\dagger(p){}_k &=& v^\dagger_-(p){}^i_k e^{-i p m} \hat a_m{}_i ,
\end{eqnarray}
\begin{eqnarray}
\hat a_n^\dagger{}^i  H_{- nm}{}_i^j \hat a_m{}_j
&=&
\int \frac{d^2 p}{(2\pi)^2} \sum_k \,
\left[
\left( \hat a_n^\dagger{}^i e^{i p n } u_-(p){}_i^k \right)
\lambda_-(p)
\left( u^\dagger_-(p){}^j_k e^{-i p m} \hat a_m{}_j \right)
\right.
\nonumber\\
&& \qquad\qquad\qquad
\left.
-\left( \hat a_n^\dagger{}^i e^{i p n } v_-(p){}_i^k \right)
 \lambda_-(p)
 \left( v^\dagger_-(p){}^j_k e^{-i p m} \hat a_m{}_j \right)
        \right]
\nonumber\\
&=&
\int \frac{d^2 p}{(2\pi)^2} \sum_k \,
        \left[
          \hat b_-^\dagger(p){}^k \lambda_-(p) \hat b_-(p){}_k
       -  \hat d_-(p){}^k \lambda_-(p) \hat d_-^\dagger(p){}_k
        \right] .
\nonumber\\
\end{eqnarray}

These creation and annihilation operators satisfy the following
commutation relations.
\begin{eqnarray}
\left\{ \hat b_+(p){}_k, \hat b_+^\dagger(q){}^l \right\}
&=& (2\pi)^2 \delta^2(p-q) \delta_k^l ,  \\
\left\{ \hat d_+(p){}^k, \hat d_+^\dagger(q){}_l \right\}
&=& (2\pi)^2 \delta^2(p-q) \delta^k_l ,  \\
\left\{ \hat b_+(p){}_k, \hat d_+(q){}^l \right\}
&=& 0 ,  \\
\left\{ \hat b_+^\dagger(p){}^k , \hat d_+^\dagger(q){}_l \right\}
&=& 0 .
\end{eqnarray}

\begin{eqnarray}
\left\{ \hat b_-(p){}_k, \hat b_-^\dagger(q){}^l \right\}
&=& (2\pi)^2 \delta^2(p-q) \delta_k^l ,  \\
\left\{ \hat d_-(p){}^k, \hat d_-^\dagger(q){}_l \right\}
&=& (2\pi)^2 \delta^2(p-q) \delta^k_l ,  \\
\left\{ \hat b_-(p){}_k, \hat d_-(q){}^l \right\}
&=& 0 ,  \\
\left\{ \hat b_-^\dagger(p){}^k , \hat d_-^\dagger(q){}_l \right\}
&=& 0 .
\end{eqnarray}

\begin{eqnarray}
\left\{ \hat b_-(p){}_k, \hat b_+^\dagger(q){}^l \right\}
&=& (2\pi)^2 \delta^2(p-q) u^\dagger_-(p) u_+(p) \delta_k^l ,  \\
\left\{ \hat b^\dagger_-(p){}^k, \hat b_+(q){}_l \right\}
&=& (2\pi)^2 \delta^2(p-q) u^\dagger_+(p) u_-(p) \delta^k_l ,  \\
\left\{ \hat d_-(p){}^k, \hat d_+^\dagger(q){}_l \right\}
&=& (2\pi)^2 \delta^2(p-q) v^\dagger_+(p)v_-(p) \delta^k_l ,  \\
\left\{ \hat d^\dagger_-(p){}_k, \hat d_+(q){}^l \right\}
&=& (2\pi)^2 \delta^2(p-q) v^\dagger_-(p) v_+(p) \delta_k^l ,  \\
\left\{ \hat b_-(p){}_k, \hat d_+(q){}^l \right\}
&=&  (2\pi)^2 \delta^2(p-q)  u^\dagger_-(p)v_+(p)\delta_k^l ,  \\
\left\{ \hat b_-^\dagger(p){}^k , \hat d_+^\dagger(q){}_l \right\}
&=& (2\pi)^2 \delta^2(p-q) v^\dagger_+(p) u_-(p) \delta^k_l ,  \\
\left\{ \hat b_+(p){}_k, \hat d_-(q){}^l \right\}
&=& (2\pi)^2 \delta^2(p-q) u^\dagger_+(p) v_-(p) \delta_k^l ,  \\
\left\{ \hat b_+^\dagger(p){}^k , \hat d_-^\dagger(q){}_l \right\}
&=& (2\pi)^2 \delta^2(p-q) v^\dagger_-(p) u_+(p) \delta^k_l.
\end{eqnarray}

Vacua of the second-quantized Hamiltonians are given by
  \begin{equation}
    \rvacp = \prod_{p,k} \hat d_+(p){}^k \rvac ,
  \end{equation}
  \begin{equation}
    \rvacm = \prod_{p,k} \hat d_-(p){}^k \rvac .
  \end{equation}

The defining creation and annihilation operators are expanded
by the creation and annihilation operators in the diagonalized bases
as follows:
\begin{eqnarray}
\hat a_n{}_i
&=&
\int \frac{d^2 p}{(2\pi)^2} e^{i p n}
\left[ u_+(p) \hat b_+(p){}_i
     + v_+(p) \hat d^\dagger_+(p){}_i \right] \\
&=&
\int \frac{d^2 p}{(2\pi)^2} e^{i p n}
\left[ u_-(p) \hat b_-(p){}_i
     + v_-(p) \hat d^\dagger_-(p){}_i \right] ,\\
\hat a^\dagger_n{}^i
&=&
\int \frac{d^2 p}{(2\pi)^2} e^{- i p n}
\left[ u^\dagger_+(p) \hat b^\dagger_+(p){}^i
      +v^\dagger_+(p) \hat d_+(p){}^i \right] \\
&=&
\int \frac{d^2 p}{(2\pi)^2} e^{- i p n}
\left[ u^\dagger_-(p) \hat b^\dagger_-(p){}^i
     + v^\dagger_-(p) \hat d_-(p){}^i \right] .
\end{eqnarray}

\section{Vacua in pure gauge limit}
\label{appendix:vacua-in-pure-gauge-limit}
\reseteqnum

In this appendix, we give the vacuum of the second-quantized
Hamiltonian theory of the three-dimensional Wilson fermion
with negative mass in the pure gauge limit.
The positive mass case can be formulated in a similar manner.
In the pure gauge limit, the interacting second-quantized
Hamiltonian with the negative mass can be written as follows:
\begin{eqnarray}
  \hat H_+ \left[g^{}_ng^\dagger_{n+\hat\mu}\right]
  =\hat G \hat H_+ \hat G^\dagger,
\quad (\beta=0),
\end{eqnarray}
where $\hat G$ is the operator of gauge transformation defined by
\begin{equation}
\hat G
= \exp \left( \sum_n \hat a_n^{\dagger i}
\{\log g^{}_n \}_i{}^j \hat a_{n j} \right) .
\end{equation}
Accordingly, the vacuum in the pure gauge limit can be expressed as
  \begin{equation}
    \hat G \rvacp .
  \end{equation}

The Hamiltonian in the pure gauge limit can be diagonalized
by the following creation and annihilation operators.

\begin{equation}
    \hat G \rvacp = \prod_{p,k} \hat d{}^G_+(p){}^k  \rvac .
\end{equation}

\begin{eqnarray}
\hat b^G_+(p){}_k
&=& u^\dagger_+(p){}_k^i e^{-i p m}
\left( g_m^\dagger{}_i^j \right)  \hat a_m{}_j .  \\
\hat b_+^{G\dagger}(p){}^k
&=& \hat a_n^\dagger{}^i
\left( g_n{}_i^j \right) e^{i p n } u_+(p){}_j^k . \\
\hat d^G_+(p){}^k
&=& \hat a_n^\dagger{}^i
\left( g_n{}_i^j \right) e^{i p n } v_+(p){}_j^k . \\
\hat d_+^{G\dagger}(p){}_k
&=& v^\dagger_+(p){}^i_k e^{-i p m}
 \left( g_m^\dagger{}_i^j \right)  \hat a_m{}_j .
\end{eqnarray}

They satisfy the following commutation relations.
\begin{eqnarray}
\left\{ \hat b^G_+(p){}_k, \hat b_+^{G\dagger}(q){}^l \right\}
&=& (2\pi)^2 \delta^2(p-q) \delta_k^l .  \\
\left\{ \hat d^G_+(p){}^k, \hat d_+^{G\dagger}(q){}_l \right\}
&=& (2\pi)^2 \delta^2(p-q) \delta^k_l .  \\
\left\{ \hat b^G_+(p){}_k, \hat d^G_+(q){}^l \right\}
&=& 0 .  \\
\left\{ \hat b_+^{G\dagger}(p){}^k , \hat d_+^{G\dagger}(q){}_l \right\}
&=& 0 .
\end{eqnarray}

\begin{eqnarray}
\left\{ \hat b_+(p){}_k, \hat b_+^{G\dagger}(q){}^l \right\}
&=& u^\dagger_+(p){}_k^i e^{-i p m}
\left( g_m{}_i^j \right) e^{i q m } u_+(q){}_j^l . \\
\left\{ \hat d_+(p){}^k, \hat d_+^{G\dagger}(q){}_l \right\}
&=&
v^\dagger_+(q){}^i_l e^{-i q m}
 \left( g_m^\dagger{}_i^j \right)
e^{i p m } v_+(p){}_j^k . \\
\left\{ \hat b_+(p){}_k, \hat d^G_+(q){}^l \right\}
&=& u^\dagger_+(p){}_k^i e^{-i p m}
\left( g_m{}_i^j \right) e^{i q m } v_+(q){}_j^l .  \\
\left\{ \hat b^G_+(p){}_k, \hat d_+(q){}^l \right\}
&=&
u^\dagger_+(p){}_k^i e^{-i p m}
\left( g_m^\dagger{}_i^j \right)  e^{i q m } v_+(q){}_j^l .  \\
\left\{ \hat b_+^{\dagger}(p){}^k , \hat d_+^{G\dagger}(q){}_l \right\}
&=&
v^\dagger_+(q){}^i_l e^{-i q m}
 \left( g_m^\dagger{}_i^j \right)  e^{i p m } u_+(p){}_j^k .\\
\left\{ \hat b_+^{G\dagger}(p){}^k , \hat d_+^{\dagger}(q){}_l \right\}
&=&
v^\dagger_+(q){}^i_l e^{-i p m}
\left( g_m{}_i^j \right) e^{i p m } u_+(p){}_j^k  .
\end{eqnarray}

In this diagonalizing basis in the pure gauge limit,
the defining creation and annihilation operators are expanded as
\begin{eqnarray}
\hat a_n{}_i
&=&
\int \frac{d^2 p}{(2\pi)^2}
\left( g_n{}_i^k \right) e^{i p n}
\left[ u_+(p) \hat b^G_+(p){}_k
    + v_+(p)  \hat d^{G\dagger}_+(p){}_k
\right] , \\
\hat a^\dagger_n{}^i
&=&
\int \frac{d^2 p}{(2\pi)^2}
\left[ \hat b^{G\dagger}_+(p){}^k u^\dagger_+(p)
      +\hat d^G_+(p){}^k  v^\dagger_+(p) \right]  e^{- i p n}
\left( g^\dagger_n{}_k^i \right) . \\
\end{eqnarray}

A similar formulation is possible for the
case of the Hamiltonian with positive mass.

\section{Action of Yukawa Coupling Operator}
\label{appendix:boundary-Yukawa-coupling}
\reseteqnum

In this appendix, we give the definition of the operator of
the boundary Yukawa coupling and its action.
The Yukawa coupling at the waveguide
boundary\cite{waveguide-analysis-golterman}
can be expressed by the following
operator\cite{boundary-yukawa-coupling-operator}
inserted
in the overlaps which implements the Wigner-Brillouin phase convention.
\begin{eqnarray}
 \hat Y
&=&
\exp \left(
\sum_{n,i} \ln y \, ( \hat a_{n 1}^\dagger{}^i \hat a_{n 1}{}_i
                    - \hat a_{n 2}^\dagger{}^i \hat a_{n 2}{}_i )
     \right)
\nonumber\\
&=&
 \prod_{n i}
\left(   \hat a_{n 1}{}_i \hat a_{n 1}^\dagger{}^i
     + y \hat a_{n 1}^\dagger{}^i \hat a_{n 1}{}_i  \right)
\left( \hat a_{n 2}{}_i \hat a_{n 2}^\dagger{}^i
     + \frac{1}{y} \hat a_{n 2}^\dagger{}^i \hat a_{n 2}{}_i \right) .
\end{eqnarray}
This Yukawa coupling operator acts on the vacua in the following manner:
  \begin{eqnarray}
\hat Y \rvacp
&=& \prod_{p,k}
\left( \hat a_n^\dagger{}^i e^{i p n } Y v_+(p){}_i^k \right) \rvac
\nonumber\\
&=& \prod_{p,k}
\left( \hat b^\dagger_+(p){}^k u^\dagger_+(p) Y v_+(p)
      +\hat d_+(p){}^k         v^\dagger_+(p) Y v_+(p)  \right) \rvac ,
  \end{eqnarray}
  \begin{eqnarray}
\hat Y \rvacm &=&
\prod_{p,k}
\left( \hat a_n^\dagger{}^i e^{i p n } Y v_-(p){}_i^k \right) \rvac
\nonumber\\
&=& \prod_{p,k}
\left( \hat b^\dagger_-(p){}^k u^\dagger_-(p) Y v_-(p)
      +\hat d_-(p){}^k         v^\dagger_-(p) Y v_-(p)  \right)\rvac ,
\end{eqnarray}
where $Y$ is the matrix in the spinor space given by
\begin{equation}
    Y = \left(
       \begin{array}{cc} y &  0 \\ 0 & \frac{1}{y} \end{array}
        \right)
      = y P_R + \frac{1}{y} P_L .
\end{equation}

\section{Calculation of the null Wess-Zumino Action}
\label{appendix:null-Wess-Zumino-action}
\reseteqnum

In this appendix, we describe the perturbative calculation of the
null Wess-Zumino action in detail. The defining overlaps
of the representation $r$ which implement the Wigner-Brillouin
phase convention can be written explicitly with the eigenvectors
as follows:
\begin{eqnarray}
\lvacm \vert \hat G^\dagger \hat Y \rvacm_r
&=&
\det_{(p,i)(q,j)}
\left[ \sum_{m}
v^\dagger_-(p) e^{-i p m}
\left( g_m^\dagger{}_i^j \right)  e^{i q m } Y v_-(q)
\right] ,
\\
\lvacp \vert \hat Y \hat G \rvacp_r
&=&
\det_{(p,i)(q,j)}
\left[ \sum_{m}
v^\dagger_+(p) Y e^{-i p m}
\left( g_m{}_i^j \right)  e^{i q m }  v_+(q)
\right] .
\end{eqnarray}
Then the explicit formula of the contribution from the
fermion in the representation $r$ to the null Wess-Zumino action
is given as
\begin{eqnarray}
i\Delta\Gamma_{WZ}[g;Y]_r
&=&
i {\rm Im} {\rm Tr} \, {\rm Ln} \left[
\sum_{m}
v^\dagger_-(p) e^{-i p m}
\left( g_m^\dagger{}_i^j \right)  e^{i q m } Y v_-(q)
\right]
\nonumber\\
&+&
i {\rm Im} {\rm Tr} \, {\rm Ln} \left[
\sum_{m}
v^\dagger_+(p) Y e^{-i p m}
\left( g_m{}_i^j \right)  e^{i q m }  v_+(q)
\right] .
\nonumber\\
\end{eqnarray}

In terms of the fluctuation of the gauge freedom,
\begin{equation}
g_m^\dagger{}_i^j  -\bbone{}_i^j
= -i  \lambda \pi{}_i^j
  +(-i)^2 \frac{1}{2!} \lambda^2 \pi^2{}_i^j
  +(-i)^3 \frac{1}{3!} \lambda^3 \pi^3{}_i^j + \cdots ,
\end{equation}
\begin{equation}
  \pi{}_i^j \equiv \pi^a  T^a{}_i^k .
\end{equation}
the defining overlap with negative mass can be expanded as follows:
\begin{eqnarray}
&&
{\rm Tr} \, {\rm Ln} \left[
\sum_{m}
v^\dagger_-(p) e^{-i p m}
\left( g_m^\dagger{}_i^j \right)  e^{i q m } Y v_-(q) \right]
\nonumber\\
&&
=
{\rm Tr} \, {\rm Ln} \left[
v^\dagger_-(p)Y v_-(p) (2\pi)^2 \delta^2(p-q)
\left( \bbone{}_i^j \right)
\right.
\nonumber\\
&& \qquad \qquad
\left.
+\sum_{m}
v^\dagger_-(p) e^{-i p m}
\left( g_m^\dagger{}_i^j  -\bbone{}_i^j \right)  e^{i q m } Y v_-(q)
\right]
\nonumber\\
&&
=
{\rm Tr} \, {\rm Ln} \left[
v^\dagger_-(p)Y v_-(p) (2\pi)^2 \delta^2(p-q)\left( \bbone{}_i^j \right)
\right]
\nonumber\\
&&
+{\rm Tr} \, {\rm Ln} \left[
(2\pi)^2 \delta^2(p-q)\left( \bbone{}_i^j \right)
\right.
\nonumber\\
&& \qquad \qquad
\left.
+\frac{1}{v^\dagger_-(p)Y v_-(p)}
\sum_{m}
v^\dagger_-(p) e^{-i p m}
\left( g_m^\dagger{}_i^j  -\bbone{}_i^j \right)  e^{i q m } Y v_-(q)
\right]
\nonumber\\
&&
=
\ln \left( \lvacm \vert \hat Y \rvacm  \right)
\nonumber\\
&& +
{\rm Tr} \, \sum_{l=1}^\infty \frac{(-1)^{l-1}}{l}
\left[
\sum_{m}
e^{-i p m} \left( g_m^\dagger{}_i^j  -\bbone{}_i^j \right) e^{i q m }
S^v_-(q;Y)
\right]^l ,
\end{eqnarray}
where
\begin{equation}
  S^v_-(p;Y) \equiv \frac{1}{v^\dagger_-(p)Y v_-(p)}
                    Y v_-(p) v^\dagger_-(p) .
\end{equation}

In the first order term, $l=1$:
\begin{equation}
{\rm Tr} \,
\left[
\sum_{m}
e^{-i p m} \left( g_m^\dagger{}_i^j  -\bbone{}_i^j \right) e^{i q m }
S_-(q;Y)
\right] ,
\end{equation}
${\cal O}(\pi)$ vanishes because of ${\rm Tr}(T^a)= 0$.
${\cal O}(\pi^2)$ is irrelevant because it is real.
${\cal O}(\pi^3)$ vanishes because of bose symmetry with respect
to the three group indices in ${\rm Tr}( T^a T^b T^c)$.

In the second order term, $l=2$:
\begin{equation}
-\frac{1}{2}
{\rm Tr} \,
\left[
\sum_{m}
e^{-i p m} \left( g_m^\dagger{}_i^j  -\bbone{}_i^j \right) e^{i q m }
S_-(q;Y)
\right]^2 ,
\end{equation}
${\cal O}(\pi^2)$ is irrelevant because it is real.
${\cal O}(\pi^3)$ vanishes because of bose symmetry with respect
to two of the three group indices in ${\rm Tr}( T^a T^b T^c)$.

The third order term, $l=3$:
\begin{equation}
\frac{1}{3}
{\rm Tr} \,
\left[
\sum_{m}
e^{-i p m} \left( g_m^\dagger{}_i^j  -\bbone{}_i^j \right) e^{i q m }
S_-(q;Y)
\right]^3
\end{equation}
has a nontrivial ${\cal O}(\pi^3)$ contribution to the null
Wess-Zumino action, which is evaluated as follows:
\begin{eqnarray}
&&
i \Delta \Gamma_{WZ}[\pi;Y]_{r-}
\nonumber\\
&&
= i \frac{1}{3} \lambda^3
\int \frac{d^2 p_1}{(2\pi)^2}
\frac{d^2 p_2}{(2\pi)^2} \frac{d^2 p_3}{(2\pi)^2}
(2\pi)^2 \delta^2(p_1+p_2+p_3)
i \epsilon^{abc} \left( \pi^a(p_1) \pi^b(p_2) \pi^c(p_3) \right)
\times
\nonumber\\
&& \qquad \qquad
A_r
\int \frac{d^2 q}{(2\pi)^2}
{\rm Tr} \left[ S^v_-(q+p_1;Y)S^v_-(q;Y)S^v_-(q-p_2;Y) \right] + \cdots ,
\end{eqnarray}
where
\begin{equation}
{\rm Tr}\left( T^a T^b T^c \right)_r = i \epsilon^{abc} A_r  .
\end{equation}

For the anomaly free theory in consideration,
\begin{equation}
  \sum_r A_r = 4 \times \frac{1}{4} - 1 = 0 .
\end{equation}
The contribution of the defining overlap with positive mass
can be also evaluated in a similar manner.
Then the induced null Wess-Zumino action vanishes completely
up to ${\cal O}(\pi^3)$.  Note that this is true before
the expansion with respect to external momenta.

The contribution of each fermion to the null Wess-Zumino action
should contain the Wess-Zumino term:
\begin{eqnarray}
\int \frac{d^2 q}{(2\pi)^2}
{\rm Tr} \left[ S_-(q+p_1;Y)S_-(q;Y)S_-(q-p_2;Y) \right]
&=& - \epsilon_{\mu\nu} p_{1\mu} p_{2\nu} c_{WZ}[Y]_-  + \cdots .
\nonumber\\
\end{eqnarray}
The coefficient $c_{WZ}[Y]_-$ is given by the following integral:
\begin{eqnarray}
c_{WZ}[Y]_-
&\equiv&
\frac{1}{2!} \epsilon_{\mu\nu}
\int \frac{d^2 q}{(2\pi)^2}
{\rm Tr} \left[ \partial_\mu S^v_-(q;Y) S^v_-(q;Y)
                \partial_\nu S^v_-(q;Y)
\right] .
\nonumber\\
\end{eqnarray}

In order to evaluate this coefficient, let us introduce the
following wave function:
\begin{equation}
v_{-}(p;Y) \equiv  Y^{1/2} v_{-}(p)
           =\left( \begin{array}{c}
       y^{1/2} \left( B(p)-m_0-\lambda_- \right) \\
       y^{-1/2} (-i)\sigma^\ast_\mu C_\mu(p) \end{array} \right) ,
\end{equation}
where $\sigma_\mu= (i,1)$.
Using this wave function, the above tree-point vertex can be
rewritten as
\begin{eqnarray}
&&{\rm Tr} \left[ S_-(q+p_1;Y)S_-(q;Y)S_-(q-p_2;Y) \right]
\nonumber\\
&=&
\frac{v^\dagger_-(q+p_1;Y) v_-(q;Y) \,
      v^\dagger_-(q;Y) v_-(q-p_2;Y) \,
      v^\dagger_-(q-p_2;Y) v_-(q+p_1;Y)}
     {v^\dagger_-(q+p_1;Y) v_-(q+p_1;Y) \,
      v^\dagger_-(q;Y) v_-(q;Y)         \,
      v^\dagger_-(q-p_2;Y) v_-(q-p_2;Y)} .
\nonumber\\
\end{eqnarray}
Then $c_{WZ}[Y]_-$ is calculated as
\begin{eqnarray}
&&
c_{WZ}[Y]_-
\nonumber\\
&&=
\frac{1}{2!} \epsilon_{\mu\nu}
\int \frac{d^2 q}{(2\pi)^2}
{\rm Tr} \left[ \partial_\mu S_-(q;Y) S_-(q;Y) \partial_\nu S_-(q;Y)
\right]
\nonumber\\
&&=
\frac{1}{2!} \epsilon_{\mu\nu}
\int \frac{d^2 q}{(2\pi)^2}
\left\{
 \frac{ \partial_\nu v^\dagger_-(q;Y) \cdot \partial_\mu v_-(q;Y) }
      { \left( v^\dagger_-(q;Y) v_-(q;Y) \right) }
\right. \nonumber\\
&& \left. \qquad\qquad\qquad\qquad
-\frac{ \partial_\nu v^\dagger_-(q;Y) \cdot v_-(q;Y)
        v^\dagger_-(q;Y) \cdot \partial_\mu v_-(q;Y) }
      { \left( v^\dagger_-(q;Y) v_-(q;Y) \right)^2 }
\right\}
\nonumber\\
&&=
(-i) \int \frac{d^2 q}{(2\pi)^2}
\frac{ \left\{ (m_0-B) \partial_\mu C_\mu \partial_\nu C_\nu
                + \partial_\mu C_\mu C_\nu \partial_\nu B
                + \partial_\nu C_\nu C_\mu \partial_\mu B  \right\} }
 {\lambda_- \left[ y (\lambda_- +m_0-B)
+ \frac{1}{y}(\lambda_- -m_0+B) \right]^2} .
\nonumber\\
\end{eqnarray}

Furthermore we can show the following relation
by the straightforward calculation.
\begin{eqnarray}
c_{WZ}[Y]_-
&=& (-i) \frac{1}{2}
\int \frac{d^2 q}{(2\pi)^2} \int_{-\infty}^\infty \frac{d \omega}{(2\pi)}
\nonumber\\
&& \quad
\frac{1}{2!} \epsilon_{\mu\nu}
{\rm Tr} \left\{ S_3(q,\omega;Y) \partial_\mu S_3^{-1}(q,\omega;Y) \times
\right.
\nonumber\\
&& \left. \qquad\qquad
                 S_3(q,\omega;Y) \partial_\nu S_3^{-1}(q,\omega;Y)
                 S_3(q,\omega;Y) \partial_\omega S_3^{-1}(q,\omega;Y)
         \right\} ,
\nonumber\\
\end{eqnarray}
where
\begin{eqnarray}
\label{eq:Chiral-propagator-with-boundary-Yukawa}
&& S_3(p,\omega; Y)  \nonumber\\
&&= \frac{ \left\{
-i \gamma_5 \omega - i \gamma_\mu C_\mu(p) +
\frac{1}{2}\left[ y(B -m_0 -\lambda_-)
                  +\frac{1}{y} (B-m_0+\lambda_-) \right] \right\} }
{\omega^2 + \lambda^{\prime 2}_- } .
\nonumber\\
\end{eqnarray}
Then, following the method of calculation in
\cite{chern-simons-current}, we obtain
\begin{eqnarray}
&&c_{WZ}[Y]_- \nonumber\\
&&=  (-i) \frac{1}{2}  \frac{1}{4\pi} \sum_{k=0}^{2} (-1)^k
      \left(
      \begin{array}{c} 2 \\ k \end{array} \right)
      \frac{ m_k } { \vert m_k \vert }
\qquad m_k= \left\{
  \begin{array}{c} -y m_0  \quad (k=0) \\
                   \frac{1}{y} (2k-m_0) \quad (k\not=0) \end{array} \right.
\nonumber\\
&&=(-i) \frac{1}{4\pi}  .
\end{eqnarray}
By the similar calculation, we find that $c_{WZ}[Y]_+$ vanishes.
Thus we obtain the correct value of the Wess-Zumino term
as far as $y$ is nonzero finite.

In order to deduce the above form of the propagator,
Eq.~(\ref{eq:Chiral-propagator-with-boundary-Yukawa}),
we note the fact that the wavefunction $v_{-}(p;Y)$ can be
an eigenvector of a certain Hamiltonian which is generalized to
include the boundary Yukawa coupling:
  \begin{eqnarray}
&&v_{-}(p;Y) v^\dagger_{-}(p;Y)
\nonumber\\
&&=\frac{1}{y(\lambda_- +m_0-B)+\frac{1}{y}(\lambda_- -m_0+B)}
\left( \begin{array}{cc}
       y(\lambda_- +m_0-B) & - C \\
       - C^\dagger & \frac{1}{y} (\lambda_- -m_0+B)
       \end{array} \right)
\nonumber\\
&&\equiv \frac{1}{2} - \frac{1}{2} \frac{H^\prime_-}{\lambda^\prime_-} .
  \end{eqnarray}
Then the Hamiltonian and the eigenvalue can be determined up to
the overall factor.
\begin{eqnarray}
&& H^\prime_- (p;Y) \equiv
\gamma_5 \left\{ i \gamma_\mu C_\mu(p) +
\frac{1}{2}\left[ y(B -m_0-\lambda_-)
                  +\frac{1}{y} (B-m_0+\lambda_-) \right] \right\} ,
\nonumber\\
\\
&&  \lambda^\prime_- \equiv
\frac{1}{2}
\left[ y(\lambda_- +m_0-B)+\frac{1}{y}(\lambda_- -m_0+B)\right] .
\end{eqnarray}
A Dirac operator can be constructed from this Hamiltonian,
\begin{eqnarray}
D_3(p,\omega;Y)
&&\equiv
i \gamma_5 \omega + \gamma_5  H^\prime_-(p;Y) \nonumber\\
&&=
i \gamma_5 \omega + i \gamma_\mu C_\mu(p) +
\frac{1}{2}\left[ y(B -m_0-\lambda_-)
                  +\frac{1}{y} (B-m_0+\lambda_-) \right] .
\nonumber\\
\end{eqnarray}
This Dirac operator satisfies the following relation,
\begin{eqnarray}
&&D_3(p,\omega;Y) D^\dagger_3(p,\omega;Y)
\nonumber\\
&&=
\omega^2 + C_\mu^2(p)
+\frac{1}{4}\left[ y(B-m_0-\lambda_-)
                  +\frac{1}{y} (B-m_0+\lambda_-) \right]^2
\nonumber\\
&&=
\omega^2 + (\lambda_- +m_0-B)(\lambda_- -m_0+B)
\nonumber\\
&&
+\frac{1}{4}\left[ y^2(B -m_0 -\lambda_-)^2
                  +\frac{1}{y^2}(B-m_0+\lambda_-)^2
                  -2(B-m_0-\lambda_-)(B-m_0+\lambda_-)\right]
\nonumber\\
&&=
\omega^2
+\frac{1}{4}\left[ y^2(B-m_0-\lambda_-)^2
                  +\frac{1}{y^2}(B-m_0+\lambda_-)^2
                  +2(B-m_0-\lambda_-)(B-m_0+\lambda_-)\right]
\nonumber\\
&&= \omega^2 + \lambda^{\prime 2}_- .
\end{eqnarray}
Then we can see that
the inverse of the Dirac operator gives the propagator,
Eq.~(\ref{eq:Chiral-propagator-with-boundary-Yukawa}).

\section{Calculation of Boundary Correlation Functions}
\label{appendix:boundary-correlation-function}
\reseteqnum

In this appendix, we describe in detail the calculation of
the boundary correlation functions with the boundary Yukawa coupling.
The overlaps which implement the Wigner-Brillouin phase convention
are evaluated as
\begin{eqnarray}
\lvacm \vert \hat G^\dagger \hat Y \rvacm
&=& \lvac \vert \prod_{p,i} d^{G\dagger}_-(p){}_i
    \hat Y    \prod_{q,j} d_-(q){}^j \rvac
\nonumber\\
&=& \lvac \vert \prod_{p,i} d^{G\dagger}_-(p){}_i
\prod_{q,j}
\left( \hat b^\dagger_-(q){}^j u^\dagger_-(q) Y v_-(q)
      +\hat d_-(q){}^j         v^\dagger_-(q) Y v_-(q)  \right) \rvac
\nonumber\\
&=&
\det_{(p,i)(q,j)}
\left[ \sum_{m,k}
v^\dagger_-(p) e^{-i p m}
\left( g_m^\dagger{}_i^j \right)  e^{i q m } Y v_-(q)
\right] .
\end{eqnarray}
\begin{eqnarray}
\lvacp \vert \hat Y \hat G \rvacp
&=& \lvac \vert \prod_{p,i} d^{\dagger}_+(p){}_i
    \hat Y    \prod_{q,j} d^G_+(q){}^j \rvac
\nonumber\\
&=& \lvac \vert
\prod_{q,j}
\left( \hat b_+(q){}_i         v^\dagger_+(q) Y u_+(q)
      +\hat d^\dagger_+(q){}_i v^\dagger_+(q) Y v_+(q)  \right)
\prod_{p,j} d^G_+(p){}^j
\rvac
\nonumber\\
&=&
\det_{(p,i)(q,j)}
\left[ \sum_{m,k}
v^\dagger_+(p) Y e^{-i p m}
\left( g_m{}_i^j \right)  e^{i q m }  v_+(q)
\right] .
\end{eqnarray}
Then the calculation of the boundary correlation function
proceeds as follows:
\begin{eqnarray}
&&
\frac{\lvacm \vert \hat G^\dagger \hat a_n{}_i \hat a^\dagger_m{}^j
    \hat Y \rvacm_r}
{\lvacm \vert \hat G^\dagger \hat Y \rvacm_r }
\nonumber\\
&=&
\lvac \vert \prod_{p,k} \hat d^{G\dagger}_-(p){}_k
\int \frac{d^2 p}{(2\pi)^2}
\left( g_n{}_i^o \right) e^{i p n}
\left[ u_-(p) \hat b^G_-(p){}_o + v_-(p) \hat d^{G\dagger}_-(p){}_o \right]
\times
\nonumber\\
&&
\int \frac{d^2 q}{(2\pi)^2}
\left[ u^\dagger_-(q) \hat b^\dagger_-(q){}^j
      +v^\dagger_-(q) \hat d_-(q){}^j \right] e^{-i q m}
\times
\nonumber\\
&&
\prod_{q,l}
\left( \hat b^\dagger_-(q){}^l u^\dagger_-(q) Y v_-(q)
      +\hat d_-(q){}^l         v^\dagger_-(q) Y v_-(q)  \right) \rvac
/ \lvacm \vert \hat G^\dagger \hat Y \rvacm_r
\nonumber\\
&=&
\int \frac{d^2 p}{(2\pi)^2} \left( g_n{}_i^o \right) e^{i p n}
\int \frac{d^2 q}{(2\pi)^2} e^{-i q m}
\times
\nonumber\\
&&
\lvac \vert \prod_{p,k} \hat d^{G\dagger}_-(p){}_k
\left[ u_-(p) \hat b^G_-(p){}_o \right]
\left[ u^\dagger_-(q) \hat b^\dagger_-(q){}^j
      +v^\dagger_-(q) \hat d_-(q){}^j \right]
\times
\nonumber\\
&& \qquad
\prod_{q,l}
\left( \hat b^\dagger_-(q){}^l u^\dagger_-(q) Y v_-(q)
      +\hat d_-(q){}^l         v^\dagger_-(q) Y v_-(q)  \right) \rvac
/ \lvacm \vert \hat G^\dagger \hat Y \rvacm_r
\nonumber\\
&=&
\int \frac{d^2 p}{(2\pi)^2} \left( g_n{}_i^o \right) e^{i p n}
\int \frac{d^2 q}{(2\pi)^2} e^{-i q m}
\times
\nonumber\\
&&
\left[
u_-(p)
\left( u^\dagger_-(p) e^{-i p r}
      \left( g^\dagger_r{}_o^j \right) e^{i q r } u_-(q)
\right)
u^\dagger_-(q)
\right.
\nonumber\\
&&
+u_-(p)
\left( u^\dagger_-(p) e^{-i p r}
      \left( g_r^\dagger{}_o^j \right)  e^{i q r } v_-(q)
\right)
v^\dagger_-(q)
\nonumber\\
&&
- u_-(p) u^\dagger_-(q)
\lvac \vert \prod_{p,k} \hat d^{G\dagger}_-(p){}_k
\hat b^\dagger_-(q){}^j \hat b^G_-(p){}_o
\times
\nonumber\\
&&
\prod_{q,l}
\left( \hat b^\dagger_-(q){}^l u^\dagger_-(q) Y v_-(q)
      +\hat d_-(q){}^l         v^\dagger_-(q) Y v_-(q)  \right) \rvac
/ \lvacm \vert \hat G^\dagger \hat Y \rvacm_r
\nonumber\\
&&
- u_-(p) v^\dagger_-(q)
\lvac \vert \prod_{p,k} \hat d^{G\dagger}_-(p){}_k
\hat d_-(q){}^j
\hat b^G_-(p){}_o
\times
\nonumber\\
&&
\left.
\prod_{q,l}
\left( \hat b^\dagger_-(q){}^l u^\dagger_-(q) Y v_-(q)
      +\hat d_-(q){}^l         v^\dagger_-(q) Y v_-(q)  \right) \rvac
/ \lvacm \vert \hat G^\dagger \hat Y \rvacm_r
\right]
\nonumber\\
&=&
\int \frac{d^2 p}{(2\pi)^2} \left( g_n{}_i^o \right) e^{i p n}
\int \frac{d^2 q}{(2\pi)^2} e^{-i q m}
\times
\nonumber\\
&&
\left[
u_-(p) u^\dagger_-(p)
e^{-i p r} \left( g^\dagger_r{}_o^j \right) e^{i q r }
\right.
\nonumber\\
&&
- u_-(p)
\left( u^\dagger_-(p)
      e^{-i p r } \left( g_r^\dagger{}_o^t \right)  e^{i p_t m }
       Y v_-(p_t) \right)
\times
\nonumber\\
&&
\left.
\left[ v^\dagger_-(p_s)
e^{-i p_s r} \left( g_r^\dagger{}_s^t \right) e^{i p_t r }
       Y v_-(p_t) \right]^{-1}_{(p_t,t)(p_s,s)}
v^\dagger_-(p_s)
   e^{-i p_s r} \left( g_r^\dagger{}_s^j \right)  e^{i q r }
\right]
\nonumber\\
&=&
\int \frac{d^2 p}{(2\pi)^2} \left( g_n{}_i^o \right) e^{i p n}
\int \frac{d^2 q}{(2\pi)^2} e^{-i q m}
\times
\nonumber\\
&&
\left[
e^{-i p r } \left( g^\dagger_r{}_o^j \right) e^{i q r }
\right.
\nonumber\\
&&
- e^{-i p r } \left( g_r^\dagger{}_o^t \right)  e^{i p_t m }
       Y v_-(p_t)
\times
\nonumber\\
&&
\left.
\left[ v^\dagger_-(p_s)
e^{-i p_s r} \left( g_r^\dagger{}_s^t \right) e^{i p_t r }
       Y v_-(p_t) \right]^{-1}_{(p_t,t)(p_s,s)}
v^\dagger_-(p_s)
   e^{-i p_s r} \left( g_r^\dagger{}_s^j \right)  e^{i q r }
\right]
\nonumber\\
&=&
\delta_{nm}\delta_i^j
- S^v_-[g](n,m;Y){}_i^s \left( g_m^\dagger{}_s^j \right) ,
\end{eqnarray}
where
\begin{eqnarray}
\label{appendix:Chiral-projection-gauge-yukawa}
S_-^v[g](n,m;Y){}_i^j
&\equiv&
\int \frac{d^2 p}{(2\pi)^2} \frac{d^2 q}{(2\pi)^2} \times
\nonumber\\
&&
e^{i p n } Y v_-(p)
\left[ v^\dagger_-(q)
e^{-i q r} \left( g_r^\dagger{}_j^i \right) e^{i p r }
      Y v_-(p) \right]^{-1}_{(p,i)(q,j)}
v^\dagger_-(q) e^{-i q m} .
\nonumber\\
\end{eqnarray}

Similarly, we obtain
\begin{eqnarray}
&&    \frac{\lvacm \vert \hat a_n{}_i \hat a^\dagger_m{}^j
      \hat G^\dagger \hat Y \rvacm_r}
     {\lvacm \vert \hat G^\dagger \hat Y \rvacm_r }
=
\delta_{nm}\delta_i^j
- \left( g_n{}_i^s \right) S^v_-[g](n,m;Y){}_s^j ,
\\
&&    \frac{\lvacm \vert \hat Y \hat G^\dagger
      \hat a_n{}_i \hat a^\dagger_m{}^j
      \rvacm_r}
     {\lvacm \vert \hat G^\dagger \hat Y \rvacm_r }
=
\delta_{nm}\delta_i^j
- Y^{-1} S^v_-[g](n,m;Y){}_i^s Y \left( g_m^\dagger{}_s^j \right) ,
\\
&&    \frac{\lvacm \vert \hat Y
      \hat a_n{}_i \hat a^\dagger_m{}^j
      \hat G^\dagger  \rvacm_r}
     {\lvacm \vert \hat G^\dagger \hat Y \rvacm_r }
=
\delta_{nm}\delta_i^j
- \left( g_n{}_i^s \right) Y^{-1} S^v_-[g](n,m;Y){}_s^j Y .
\end{eqnarray}
Therefore we finally have
\begin{eqnarray}
\langle \phi_n{}_i \phi^\dagger_m{}^i  \rangle_{-r}^{IN}
&=&\frac{1}{Z} \int d\mu[g;K,Y]
\left[ \frac{1}{2} \delta_{nm} \delta_i^i
- S_-^v[g](n,m;Y){}_i^o \left( g^\dagger_m{}_o^i \right)
\right] ,
\\
\langle \varphi_n{}_i \varphi^\dagger_m{}^i  \rangle_{-r}^{IN}
&=&\frac{1}{Z} \int d\mu[g;K,Y]
\left[ \frac{1}{2} \delta_{nm} \delta_i^i
- \left( g^\dagger_m{}_i^o \right) S_-^v[g](n,m;Y){}_o^i
\right] ,
\\
\langle \phi_n{}_i \phi^\dagger_m{}^i  \rangle_{-r}^{OUT}
&=&\frac{1}{Z} \int d\mu[g;K,Y]
\left[ \frac{1}{2} \delta_{nm} \delta_i^i
- Y^{-1} S_-^v[g](n,m;Y){}_i^o Y \left( g^\dagger_m{}_o^i \right)
\right] ,
\nonumber\\
\\
\langle \varphi_n{}_i \varphi^\dagger_m{}^i  \rangle_{-r}^{OUT}
&=&\frac{1}{Z} \int d\mu[g;K,Y]
\left[ \frac{1}{2} \delta_{nm} \delta_i^i
- \left( g^\dagger_m{}_i^o \right) Y^{-1} S_-^v[g](n,m;Y){}_o^i Y
\right] .
\nonumber\\
\end{eqnarray}

\section{Expansion of $S_-^v[g](n,m;Y){}_i^j$ in $\lambda$}
\label{appendix:expansion-in-lambda}
\reseteqnum

In this appendix, $S_-^v[g;Y](n;m){}_i^j$ defined by
Eq.~(\ref{appendix:Chiral-projection-gauge-yukawa})
is evaluated in the expansion of $\lambda$ for the gauge freedom
configuration:
\begin{equation}
g_m^\dagger{}_i^j
= \bbone{}_i^j -i  \lambda \pi{}_i^j
  +(-i)^2 \frac{1}{2!} \lambda^2 \pi^2{}_i^j
  +(-i)^3 \frac{1}{3!} \lambda^3 \pi^3{}_i^j + \cdots .
\end{equation}

The matrix to be inverted in
Eq.~(\ref{appendix:Chiral-projection-gauge-yukawa})
can be expanded in $\lambda$ as
\begin{eqnarray}
&&  \left[ v^\dagger_-(p_s)
e^{-i p_s r} \left( g_r^\dagger{}_j^t \right) e^{i p_t r }
      Y v_-(p_t) \right]
\nonumber\\
&&=
\left[ v^\dagger_-(p_s) Y v_-(p_s) \right]
(2\pi)^2 \delta^2(p_s-p_t) \left( \bbone{}_j^t \right)
\nonumber\\
&&+
\left[ v^\dagger_-(p_s)
e^{-i p_s r} \left( (-i\lambda) \pi_r{}_j^t \right)  e^{i p_t r }
      Y v_-(p_t) \right]
\nonumber\\
&&+
\left[ v^\dagger_-(p_s)
e^{-i p_s r} \left(
\frac{(-i\lambda)^2}{2!}\pi_r^2{}_j^t \right)  e^{i p_t r }
      Y v_-(p_t) \right] +{\cal O}(\lambda^3) .
\nonumber\\
\end{eqnarray}
Then its inverse reads
\begin{eqnarray}
&&\left[ v^\dagger_-(p_s)
e^{-i p_s r} \left( g_r^\dagger{}_j^t \right) e^{i p_t r }
      Y v_-(p_t) \right]^{-1}_{(p_t,t)(p_s,j)}
\nonumber\\
&&=
\frac{1}{\left[ v^\dagger_-(p_t) Y v_-(p_t) \right]}
(2\pi)^2 \delta^2(p_t-p_s) \left( \bbone{}_t^j \right)
\nonumber\\
&&
-\frac{1}{\left[ v^\dagger_-(p_t) Y v_-(p_t) \right]}
\left[ v^\dagger_-(p_t)
e^{-i p_t r} \left( (-i\lambda) \pi_r{}_t^j \right)  e^{i p_s r }
      Y v_-(p_s) \right]
\frac{1}{\left[ v^\dagger_-(p_s) Y v_-(p_s) \right]}
\nonumber\\
&&-
\frac{1}{\left[ v^\dagger_-(p_t) Y v_-(p_t) \right]}
\left[ v^\dagger_-(p_t)
e^{-i p_t r} \left(
\frac{(-i\lambda)^2}{2!}\pi_r^2{}_t^j \right)  e^{i p_s r }
      Y v_-(p_s) \right]
\frac{1}{\left[ v^\dagger_-(p_s) Y v_-(p_s) \right]}
\nonumber\\
&&+
\frac{1}{\left[ v^\dagger_-(p_t) Y v_-(p_t) \right]}
\left[ v^\dagger_-(p_t)
e^{-i p_t r} \left( (-i\lambda) \pi_r{}_t^u \right)  e^{i p_u r }
      Y v_-(p_u) \right]
\times
\nonumber\\
&& \quad
\frac{1}{\left[ v^\dagger_-(p_u) Y v_-(p_u) \right]}
\left[ v^\dagger_-(p_u)
e^{-i p_u r} \left( (-i\lambda) \pi_r{}_u^j \right)  e^{i p_s r }
      Y v_-(p_s) \right]
\frac{1}{\left[ v^\dagger_-(p_s) Y v_-(p_s) \right]}
\nonumber\\
&&
+{\cal O}(\lambda^3) .
\nonumber\\
\end{eqnarray}
Therefore we obtain the following expression
\begin{eqnarray}
S_-^v[g](n,m;Y){}_i^j
&=& S_-^v(n-m;Y) \delta_i^j \nonumber\\
&-&\sum_r
   S_-^v(n-r;Y) \left( (-i\lambda) \pi_r{}_i^j \right) S_-^v(r-m;Y)
\nonumber\\
&-& \sum_r
  S_-^v(n-r;Y)
 \left( \frac{(-i\lambda)^2}{2!}\pi_r^2{}_t^j \right) S_-^v(r-m;Y)
\nonumber\\
&+&\sum_{rl} S_-^v(n-r;Y) \left( (-i\lambda) \pi_r{}_i^o \right)
             S_-^v(r-l;Y) \left( (-i\lambda) \pi_r{}_o^j \right)
             S_-^v(l-m;Y)
\nonumber\\
&+&{\cal O}(\lambda^3) .
\end{eqnarray}

\section{$S_-^v[g;Y](n;m){}_i^j$ in the limits $y=0$ and $y=\infty$}
\label{appendix:expansion-in-Y}
\reseteqnum

In this appendix, $S_-^v[g;Y](n;m){}_i^j$ defined by
Eq.~(\ref{appendix:Chiral-projection-gauge-yukawa})
is evaluated in the limits $y=0$ and $y=\infty$.

The matrix to be inverted in the above formula can
be written explicitly as
\begin{eqnarray}
&&  \left[ v^\dagger_-(q)
e^{-i q r} \left( g_r^\dagger{}_j^i \right) e^{i p r }
      Y v_-(p) \right]
\nonumber\\
&&=
y
\frac{\left(B(q)-m_0-\lambda_-(q) \right)}
     {\sqrt{2\lambda_-(q)(\lambda_-(q) + m_0-B(q))}}
   e^{-i q r} \left( g_r^\dagger{}_j^i \right) e^{i p r }
\frac{\left(B(p)-m_0-\lambda_-(p) \right)}
     {\sqrt{2\lambda_-(p)(\lambda_-(p) + m_0-B(p))}}
\nonumber\\
&&
+\frac{1}{y}
\frac{C(q)}
     {\sqrt{2\lambda_-(q)(\lambda_-(q) + m_0-B(q))}}
 e^{-i q r} \left( g_r^\dagger{}_j^i \right) e^{i p r }
\frac{C^\dagger(p)}
     {\sqrt{2\lambda_-(p)(\lambda_-(p) + m_0-B(p))}} .
\nonumber\\
\end{eqnarray}
If $y << 1$, we may rewrite the expression as
\begin{eqnarray}
&&  \left[ v^\dagger_-(q)
e^{-i q r} \left( g_r^\dagger{}_j^i \right) e^{i p r }
      Y v_-(p) \right]
\nonumber\\
&&=
\frac{1}{\sqrt{2\lambda_-(q)(\lambda_-(q) + m_0-B(q))}
         \sqrt{2\lambda_-(p)(\lambda_-(p) + m_0-B(p))}}
\times
\nonumber\\
&&
\frac{1}{y}
C(q)
e^{-i q r} \left( g_r^\dagger{}_j^s \right) e^{i p_s r }
C^\dagger(p_s)
\times
\nonumber\\
&&
\left\{
(2\pi)^2 \delta(p_s-p) \delta_s^i
\phantom{\frac{1}{C^\dagger(p_s)}}
\right.
\nonumber\\
&&
+ y^2
\frac{1}{C^\dagger(p_s)}
   e^{-i p_s r} \left( g_r{}_s^t \right) e^{i p_t r } \times
\nonumber\\
&&
\left. \quad
\frac{\left(B(p_t)-m_0-\lambda_-(p_t) \right)}
     {C(p_t)}
   e^{-i p_t l} \left( g_{l}^\dagger{}_t^i \right) e^{i p l }
    \left(B(p)-m_0-\lambda_-(p) \right)
\right\} .
\nonumber\\
\end{eqnarray}
The inverse of this matrix then reads
\begin{eqnarray}
&&  \left[ v^\dagger_-(q)
e^{-i q r} \left( g_r^\dagger{}_j^i \right) e^{i p r }
      Y v_-(p) \right]^{-1} (p,i)(q,j)
\nonumber\\
&&=
\sqrt{2\lambda_-(p)(\lambda_-(p) + m_0-B(p))}
\sqrt{2\lambda_-(q)(\lambda_-(q) + m_0-B(q))}
\times
\nonumber\\
&&
\left\{
(2\pi)^2 \delta(p_s-p) \delta_s^i
\phantom{\frac{1}{C^\dagger(p_s)}}
\right.
\nonumber\\
&& \qquad
+ y^2
\frac{1}{C^\dagger(p_s)} \,
   e^{-i p_s r} \left( g_r{}_s^t \right) e^{i p_t r }
\frac{\left(B(p_t)-m_0-\lambda_-(p_t) \right)}
     {C(p_t)} \times
\nonumber\\
&& \left.
\quad\qquad\qquad\phantom{\frac{1}{C^\dagger(p_s)}}
   e^{-i p_t l} \left( g_{l}^\dagger{}_t^i \right) e^{i p l }
    \left(B(p)-m_0-\lambda_-(p) \right) \,
\right\}^{-1}_{(p,i)(p_s,s)}
\times
\nonumber\\
&&
y
\frac{1}{C^\dagger(p_s)}
e^{-i p_s r} \left( g_r{}_s^j \right) e^{i q r }
\frac{1}{C(q)} .
\end{eqnarray}
Then the expression ready for the expansion in $y$ is given by
\begin{eqnarray}
&&S_-^v[g](n,m;Y){}_i^j
\nonumber\\
&&=
\int \frac{d^2 p}{(2\pi)^2} \frac{d^2 q}{(2\pi)^2}
\times
\nonumber\\
&&
e^{i p n }
\left( \begin{array}{c}
          y^2 \left( B(p)-m_0-\lambda_-(p) \right) \\
          C^\dagger(p)
       \end{array} \right)
\left( \begin{array}{cc}
    B(q)-m_0-\lambda_-(q) & C(q)
       \end{array} \right) e^{-i q m}
\times
\nonumber\\
&&
\left\{
(2\pi)^2 \delta(p_s-p) \delta_s^i
\phantom{\frac{1}{C^\dagger(p_s)}}
\right.
\nonumber\\
&& \quad
+ y^2
\frac{1}{C^\dagger(p_s)} \,
   e^{-i p_s r} \left( g_r{}_s^t \right) e^{i p_t r }
\frac{\left(B(p_t)-m_0-\lambda_-(p_t) \right)}
     {C(p_t)} \times
\nonumber\\
&& \left.
\quad\qquad\qquad\phantom{\frac{1}{C^\dagger(p_s)}}
   e^{-i p_t l} \left( g_{l}^\dagger{}_t^i \right) e^{i q l }
    \left(B(p)-m_0-\lambda_-(p) \right) \,
\right\}^{-1}_{(p,i)(p_s,s)}
\times
\nonumber\\
&& \quad
\frac{1}{C^\dagger(p_s)}
e^{-i p_s r} \left( g_r{}_s^j \right) e^{i q r }
\frac{1}{C(q)} .
\end{eqnarray}

On the other hand, if $y >> 1$, we may rewrite the expression as
\begin{eqnarray}
&&  \left[ v^\dagger_-(q)
e^{-i q r} \left( g_r^\dagger{}_j^i \right) e^{i p r }
      Y v_-(p) \right]
\nonumber\\
&&=
\frac{1}{\sqrt{2\lambda_-(q)(\lambda_-(q) + m_0-B(q))}
         \sqrt{2\lambda_-(p)(\lambda_-(p) + m_0-B(p))}}
\times
\nonumber\\
&&
y
\left(B(q)-m_0-\lambda_-(q) \right)
e^{-i q r} \left( g_r^\dagger{}_j^s \right) e^{i p r }
\left(B(p_s)-m_0-\lambda_-(p_s) \right)
\times
\nonumber\\
&&
\left\{
(2\pi)^2 \delta(p_s-p) \delta_s^i
\phantom{\frac{1}{C^\dagger(p_s)}}
\right.
\nonumber\\
&&
+ \frac{1}{y^2}
\frac{1}{\left(B(p_s)-m_0-\lambda_-(p_s) \right)}
   e^{-i p_s r} \left( g_r{}_s^t \right) e^{i p_t r } \times
\nonumber\\
&&
\left. \qquad\qquad\qquad
\frac{C(p_t)}
     {\left(B(p_t)-m_0-\lambda_-(p_t) \right)}
   e^{-i p_t l} \left( g_{l}^\dagger{}_t^i \right) e^{i p l }
    C^\dagger(p)
\right\} .
\nonumber\\
\end{eqnarray}
The inverse of this matrix then reads
\begin{eqnarray}
&&  \left[ v^\dagger_-(q)
e^{-i q r} \left( g_r^\dagger{}_j^i \right) e^{i p r }
      Y v_-(p) \right]^{-1} (p,i)(q,j)
\nonumber\\
&&=
\sqrt{2\lambda_-(p)(\lambda_-(p) + m_0-B(p))}
\sqrt{2\lambda_-(q)(\lambda_-(q) + m_0-B(q))}
\times
\nonumber\\
&&
\left\{
(2\pi)^2 \delta(p_s-p) \delta_s^i
\phantom{\frac{1}{C^\dagger(p_s)}}
\right.
\nonumber\\
&&
+ \frac{1}{y^2}
\frac{1}{\left(B(p_s)-m_0-\lambda_-(p_s) \right)}
   e^{-i p_s r} \left( g_r{}_s^t \right) e^{i p_t r } \times
\nonumber\\
&& \left.
\qquad
\frac{C(p_t)}
     {\left(B(p_t)-m_0-\lambda_-(p_t) \right)}
   e^{-i p_t l} \left( g_{l}^\dagger{}_t^i \right) e^{i p l }
    C^\dagger(p)
\right\}^{-1}_{(p,i)(p_s,s)}
\times
\nonumber\\
&&
\frac{1}{y}
\left(B(p_s)-m_0-\lambda_-(p_s) \right)
e^{-i p_s r} \left( g_r^\dagger{}_j^s \right) e^{i q r }
\left(B(q)-m_0-\lambda_-(q) \right) .
\end{eqnarray}
Then the expression ready for the expansion in $1/y$ is given by
\begin{eqnarray}
&&S_-^v[g](n,m;Y){}_i^j
\nonumber\\
&&=
\int \frac{d^2 p}{(2\pi)^2} \frac{d^2 q}{(2\pi)^2}
\times
\nonumber\\
&&
e^{i p n }
\left( \begin{array}{c}
          B(p)-m_0-\lambda_-(p) \\
          \frac{1}{y^2} C^\dagger(p)
       \end{array} \right)
\left( \begin{array}{cc}
    B(q)-m_0-\lambda_-(q) & C(q)
       \end{array} \right) e^{-i q m}
\times
\nonumber\\
&&
\left\{
(2\pi)^2 \delta(p_s-p) \delta_s^i
\phantom{\frac{1}{C^\dagger(p_s)}}
\right.
\nonumber\\
&&
+ \frac{1}{y^2}
\frac{1}{\left(B(p_s)-m_0-\lambda_-(p_s) \right)}
   e^{-i p_s r} \left( g_r{}_s^t \right) e^{i p_t r } \times
\nonumber\\
&& \left.
\qquad
\frac{C(p_t)}
     {\left(B(p_t)-m_0-\lambda_-(p_t) \right)}
   e^{-i p_t l} \left( g_{l}^\dagger{}_t^i \right) e^{i p l }
    C^\dagger(p)
\right\}^{-1}_{(p,i)(p_s,s)}
\times
\nonumber\\
&&
\frac{1}{\left(B(p_s)-m_0-\lambda_-(p_s) \right)}
e^{-i p_s r} \left( g_r{}_j^s \right) e^{i q r }
\frac{1}{\left(B(q)-m_0-\lambda_-(q) \right)} .
\end{eqnarray}

\end{document}